\documentclass[lettersize,journal]{IEEEtran}
\usepackage{graphicx}
\usepackage{amsmath,amsfonts}
\usepackage{algorithmic}
\usepackage{algorithm}
\usepackage{array}
\usepackage[caption=false,font=normalsize,labelfont=sf,textfont=sf]{subfig}
\usepackage{textcomp}
\usepackage{stfloats}
\usepackage{url}
\usepackage{verbatim}
\usepackage{graphicx}
\usepackage{cite}
\usepackage{amssymb}
\usepackage{booktabs}
\usepackage{xcolor}
\usepackage{tabularx}   
\begin{document}

\title{Energy-Based Cell Association in Nonuniform Renewable Energy-Powered Cellular Networks: Analysis and Optimization of Carbon Efficiency}

\author{Yuxi Zhao,~\IEEEmembership{Student Member,~IEEE}, 
Vicente Casares-Giner,~\IEEEmembership{Life Member,~IEEE}, 
Vicent Pla, 
Luis Guijarro, 
Iztok Humar,~\IEEEmembership{Senior Member,~IEEE}, 
Yi Zhong,~\IEEEmembership{Senior Member,~IEEE}, 
Xiaohu Ge,~\IEEEmembership{Senior Member,~IEEE}%
\thanks{This research is supported in part by the National Key R\&D Program of China under Grant 2024YFE0200504. \textit{(Corresponding authors: Yi Zhong.)}}
\thanks{Y.~Zhao, Y.~Zhong, and X.~Ge are with the School of Electronic Information and Communications, Huazhong University of Science and Technology, Wuhan, Hubei 430074, China  (e-mail: zhao\_yuxi, yzhong@hust.edu.cn, xhge@mail.hust.edu.cn).}%
\thanks{V.~Casares-Giner, V.~Pla, and L.~Guijarro are with the 
  Universitat Polit\`ecnica de Val\`encia, Spain.
  Part of this work was conducted while these authors were visiting Huazhong University of Science and Technology, Wuhan, P.R.~China. 
  This visit was funded by the Ministry of Science and Technology (MOST) of P.R.~China
  through project \mbox{no. 1 10000206520238034}.
  The work of these authors was supported in part by Grant PID2021-123168NB-I00, 
  funded by MCIN/AEI, Spain/10.13039/50 1100011033 
  and the European Union A way of making Europe/ERDF, 
  and Grant TED2021-131387B-I00, funded by MCIN/AEI, 
  Spain/10.13039/501100011033 and the European Union NextGenerationEU/RTRP. (e-mail: vcasares@dcom.upv.es, vpla@upv.es, lguijar@dcom.upv.es)
}
\thanks{I. Humar is with the Faculty of Electrical Engineering, University of Ljubljana, Slovenia. His work is supported in part by the Slovenian Research and Innovation Agency program P2-0425.(e-mail:iztok.humar@fe.uni-lj.si)}%
}



\maketitle

\begin{abstract}
The increasing global push for carbon reduction highlights the importance of integrating renewable energy into the supply chain of cellular networks. 
However, due to the stochastic nature of renewable energy generation and the uneven load distribution across base stations, the utilization rate of renewable energy remains low. 
To address these challenges, this paper investigates the trade-off between carbon emissions and downlink throughput in cellular networks, offering insights into optimizing both network performance and sustainability. The renewable energy state of base station batteries and the number of occupied channels are modeled as a quasi-birth-death process. We construct models for the probability of channel blocking, average successful transmission probability for users, downlink throughput, carbon emissions, and carbon efficiency based on stochastic geometry. Based on these analyses, an energy-based cell association scheme is proposed to optimize the carbon efficiency of cellular networks. 
The results show that, compared to the closest cell association scheme, the energy-based cell association scheme is capable of reducing the carbon emissions of the network by 13.0\% and improving the carbon efficiency by 11.3\%.
\end{abstract}


\section{Introduction}
\subsection{Background and Motivation}
\IEEEPARstart{I}{n} 2020, international organizations such as the International Telecommunication Union and the Global System for Mobile Communications Association issued a call to action to the global information and communications technology (ICT) industry, urging a reduction in carbon emissions (CEm) of 45\% by 2030 \cite{l_itu-t_0120_greenhouse_nodate}. Nevertheless, the advent of 5G has precipitated a swift surge in CEm within the ICT industry, together with a notable deterioration in carbon efficiency (CEf) \cite{li_carbon_2023}. Based on current projections, the ICT sector's CEm are expected to reach 125 million tons of $\text{CO}_2$ by 2030, representing 1.97\% of global emissions \cite{zhou_how_2019}. This is due to the fact that, as ICT continues to experience exponential growth in energy consumption, traditional fossil fuels remain the primary source of ICT energy. So, it is imperative to use renewable energy (RE), such as solar and wind energy, to supply ICT equipment. 

However, the generation of RE is random and uneven \cite{khalyasmaa_prediction_2019}, which can be easily affected by weather and geographic location. On the other hand, the distribution of cellular-network loads is uneven, with concentrations occurring at specific locations \cite{cici_decomposition_2015,niu_tango_2011}. These locations, commonly referred to as "social hot spots," are areas where users tend to congregate, such as commercial areas during the day and residential areas at night. The traditional cell association scheme is aim at improving the signal-to-interference-plus-noise ratio (SINR) and throughput, so users always associate with the base stations (BSs) offering the strongest signal \cite{mei2021performance, si2021qos}. Given the unequal distribution of users, these association schemes can result in an unbalanced load on BSs. Specifically, the BSs in social hot spot areas are fully loaded, while those in other areas are either idle or lightly loaded. Energy consumption has a positive correlation with the load of a BS~\cite{auer_how_2011}, leading to uneven energy consumption across BSs in the network when traditional association schemes are employed. A misalignment exists between the energy storage of RE and the energy consumption of BSs: BSs with low or no load often have a surplus of RE, while those with high load experience a shortage. This leads to the consumption of grid energy, predominantly generated from non-renewable sources, resulting in CEm. The low utilization rate of RE further hinders efforts to reduce the network's CEm.

\subsection{Related Works}

\begin{table*}[tbh]
\centering
\caption{Related Works}\label{tab1}
\renewcommand{\arraystretch}{1.5} 
\setlength{\tabcolsep}{12pt} 
\begin{tabularx}{\linewidth}{
    >{\raggedright\bfseries}X@{} 
    >{\raggedright\arraybackslash}X
    >{\raggedright\arraybackslash}X
}
\toprule[1.5pt]
\multicolumn{1}{>{\centering\arraybackslash}p{2.5cm}}{\textbf{Research area}} & 
\multicolumn{1}{>{\centering\arraybackslash}p{4.5cm}}{\textbf{Reference}} & 
\multicolumn{1}{>{\centering\arraybackslash}p{4cm}}{\textbf{Limitations}} \\
\midrule[1pt]

\textbf{Network performance analysis}
& \cite{liu_performance_2018,huang2015renewable}: use stochastic geometry to analyze the network performance,\newline
  \cite{zhao2024carbon,dhillon_fundamentals_2014,parzysz_power-availability-aware_2017,yu_traffic_2016,lam_energy_2020,li_stochastic_2015,ghazanfari2016ambient,khan_downlink_2016,chu2021stochastic}: use a Markov chain to model the network performance
& The one-dimensional Markov chain was used and the influence of the number of occupied channels was overlooked \\ 
\addlinespace[6pt]

\midrule[0.8pt]

\textbf{The distribution of users and BSs} 
& \cite{wang2024performance,afshang_poisson_2018,saha2019unified,saha_rate_2020,li_downlink_2017,wang_effect_2020}: Poisson cluster process (PCP) distribution.\newline
  \cite{ullah2020performance}: Thomas cluster process (TCP)
& The performance of RE-powered cellular networks under the nonuniform-user-distribution model has not been investigated. \\ 
\addlinespace[6pt]

\midrule[0.8pt]

\textbf{Network optimization}
& \cite{han_green-energy_2013}: latency-aware cell zooming scheme.\newline
  \cite{zhang_energy-aware_2017,fletscher_energy-aware_2019,rehman2018joint,aktar2019user,liu_green_2016,qiu_power-aware_2023}: energy-aware cell association scheme.\newline
  \cite{xu2016user}: bandwidth-aware cell association scheme.
& Few works optimize CEf.
The impact of the cell association scheme on network performance has not been analyzed. \\ 

\bottomrule[1.5pt]
\end{tabularx}
\end{table*}

To solve this problem, an increasing number of studies have focused on investigating the performance and feasibility of RE-powered cellular networks. These aim to provide valuable insights into the optimization of RE-powered cellular networks. Stochastic geometry has emerged as a key analytical tool for modeling spatial distributions of BSs and users.  In \cite{liu_performance_2018} a Poisson hole process is employed to analyze coverage probabilities in ultra-dense small-cell networks with non-uniform BS deployment, highlighting the impact of renewable energy intermittency on network reliability.  An energy field model with spatial correlation was proposed in \cite{huang2015renewable} to quantify coverage degradation caused by renewable energy fluctuations, introducing energy aggregation techniques to stabilize power supply. Besides, Markov chains have been widely adopted to characterize the temporal dynamics of energy harvesting and consumption. In \cite{zhao2024carbon} a spatial-temporal model combining stochastic geometry and queuing theory was introduced to quantify the probability of coverage and carbon efficiency in RE-powered networks, highlighting the trade-off between traffic offloading and energy sharing. In \cite{dhillon_fundamentals_2014}, BS availability was modeled through a birth-death process to derive the steady-state probability of energy-sufficient BSs, while \cite{parzysz_power-availability-aware_2017} proposed a cell association strategy based on Markovian battery dynamics, which takes power availability into account. These works demonstrated how finite battery capacity and energy arrival rates influence network performance. Additionally, Discrete-time Markov chains with stochastic geometry was integrated in \cite{yu_traffic_2016}, and energy efficiency and throughput in heterogeneous networks were analyzed, revealing the interplay between energy harvesting capabilities and traffic offloading. Performance metrics such as energy efficiency, outage probability, and successful transmission rate have been rigorously analyzed in \cite{lam_energy_2020,li_stochastic_2015}. Besides, \cite{ghazanfari2016ambient} investigated ambient RF energy harvesting in ultra-dense networks, emphasizing the trade-off between harvested energy and co-channel interference.  Furthermore, cooperative transmission in large-scale networks was analyzed in \cite{khan_downlink_2016}, showing that clustered energy harvesting improves link success probability but requires careful optimization of cluster size.  In \cite{chu2021stochastic}, this framework was extended to Device to Device communication, proposing guard zone rules to balance energy harvesting opportunities and interference avoidance. In the aforementioned studies, a one-dimensional Markov chain was typically employed to model the RE state of the BSs, with the influence of the number of occupied channels being overlooked. However, the performance and the energy consumption of the BSs indeed depend on the number of occupied channels. In this paper, the RE state and the number of occupied channels together define the state of a single quasi-birth-death (QBD) process, which is used to effectively analyze the network’s performance.

Moreover, the Poisson point process (PPP) was employed in the aforementioned studies to model the uniform distribution of the BSs and the users. In light of the tendency of users to congregate in social hot spots, an increasing number of studies employ the Poisson cluster process (PCP) to model the nonuniform distribution of users\cite{wang2024performance}. Based on the PCP, the performance of the network, including the probability of coverage and the throughput, was analyzed with different cell association schemes in \cite{afshang_poisson_2018}. In a unified analytical framework, \cite{saha2019unified} addressed the coverage probability under max-power association in multi-tier networks with mixed PPP and PCP deployments. Besides, the probability mass function (PMF) of the users in a cell was derived from the PCP-based user distribution in \cite{saha_rate_2020}.  Using the PCP model, the probability of coverage was derived in \cite{li_downlink_2017}. The spatiotemporal statistics of network traffic, total arrival rate, queue stability, and delay were analyzed with PPP-based and PCP-based user distributions in \cite{wang_effect_2020}. Recent advancements have further explored user-centric deployment strategies in heterogeneous networks. A TCP-based model that integrates both clustered and uniformly distributed users was proposed in \cite{ullah2020performance}, and closed-form expressions for outage probability and rate coverage under load balancing mechanisms were derived. Nevertheless, the performance of RE-powered cellular networks under the nonuniform user distribution model, particularly a user distribution based on PCP, has yet to be investigated.

To optimize network performance, aligning the RE supply with power demand in cellular networks can be achieved by optimizing the cell association scheme. Foundational studies established critical tradeoffs. In \cite{han_green-energy_2013}, a latency-aware cell zooming scheme was proposed to balance green energy usage and service delay, while \cite{zhang_energy-aware_2017} introduced energy-biased maximum-received-power association. However, a simple model was employed for the energy bias, with no consideration given to its optimality. dual-timescale control was demonstrated in \cite{fletscher_energy-aware_2019},  which combines real-time solar-powered small cell association with predictive battery scheduling. In \cite{xu2016user}, a distributed association algorithm for RE-powered mmWave massive MIMO heterogeneous networks was designed, tripling network throughput through bandwidth-aware load distribution. A joint user association and BS switching scheme for heterogeneous networks that dynamically adjusts cell sizes and operation modes was developed in \cite{rehman2018joint} to balance harvested energy utilization with interference reduction. And in \cite{aktar2019user}, SINR-based and green-aware association policies in cloud RANs were systematically compared, revealing SINR strategies achieve superior energy efficiency through optimized signal quality. Battery management plays a critical role in renewable energy utilization. A dual-battery system with residual green energy ratio-aware association was introduced in \cite{liu_green_2016}, where BSs dynamically adjust cell sizes based on real-time energy reserves. The author of \cite{qiu_power-aware_2023} addressed battery leakage effects by developing a Lyapunov-optimized joint power-aware user association and renewable energy configuration scheme, effectively balancing traffic load and energy consumption under non-ideal storage conditions. Nevertheless, the majority of the aforementioned studies have concentrated on reducing grid energy consumption, while few ave addressed on CEf. Moreover, most existing studies propose cell association schemes without thoroughly examining their impact on the performance of RE-powered networks.

\subsection{Contributions}
This paper analyzes the impact of cell association schemes on the performance of RE-powered networks, including CEm and CEf, with nonuniformly distributed users concentrated in areas surrounding social hot spots. It is demonstrated that there is an energy-based cell association scheme that maximizes the network's CEf. Based on this result, an energy-based cell association scheme is proposed to reduce the CEm of networks and improve the CEf. The main contributions of this paper are as follows:

\begin{enumerate}
\item{In order to address the randomness of user arrivals, user departures, and RE generation in RE-powered cellular networks, the RE state and the number of occupied channels are modeled as a QBD. The channel-blocking probability, the grid energy-consumption probability, and the channel-occupancy probability of the BSs are derived.}
\item{Based on stochastic geometry, the average successful transmission probability of users, downlink throughput, CEm, CEf, and power consumption in the cellular network with a nonuniform user distribution and a biased maximum long-term received-power cell association scheme are derived. In addition, it is demonstrated that the energy-based cell association scheme can reduce the CEm of networks and that there is an energy-based cell association scheme that maximizes the network's CEf.}
\item{A novel energy-based cell association scheme, based on a genetic algorithm, is proposed. This scheme can identify the optimal cell association bias for BSs, thereby enhancing the CEf and reducing the CEm of networks, while ensuring the quality of service (QoS). Simulation results show that, compared to the conventional nearest-cell association scheme, the CEm of the network is reduced by 13.0\%, while the CEf is increased by 11.3\%.}
\end{enumerate}

\subsection{Organization}
The rest of the paper is organized as follows. Section II introduces the system model of the RE-powered cellular network with a nonuniform user distribution. The average successful transmission probability of the user and the downlink throughput are derived in Section III.
Besides, the CEm model, and the CEf model are constructed in Section III. 
Based on Section III, the impact of cell association schemes on several key performance indicators is examined in Section IV, including the average successful transmission probability of the users, the power consumption, the CEm, and the CEf. Using a genetic algorithm, an energy-based cell association scheme is proposed and evaluated in Section V. The discussion of the paper is in Section VI. Finally, Section VII concludes the paper.

\section{System model}
The downlink communication scenario of typical cellular networks is considered in this paper. The locations of BSs are modeled according to a homogeneous PPP $\Phi_B$ with density $\lambda_B$. There are two tiers of users: users with a uniform distribution and users with a nonuniform distribution. The locations of the users with a uniform distribution are modeled according to a homogeneous PPP $\Phi_{u(1)}$ with density $\lambda_{u(1)}$, while the locations of the users with a nonuniform distribution are modeled according to a PCP \cite{khan_downlink_2016}. In the PCP, the locations of cluster centers (social hotspots) are modeled according to a homogeneous PPP $\Phi_{p}$ with density $\lambda_{p}$. The users with a nonuniform distribution are distributed according to a homogeneous PPP $\Phi_{u(2)}$  with density $\lambda_{u(2)}$ within circular regions of radius $r$, situated around the cluster centers. Denote $\Phi_{u(2),k}$ represents the homogeneous PPP with density $\lambda_{u(2)}$ around the cluster center $y_k \in \Phi_{p}$.

\begin{figure}[!t]
\centering
\includegraphics[width=3.5in]{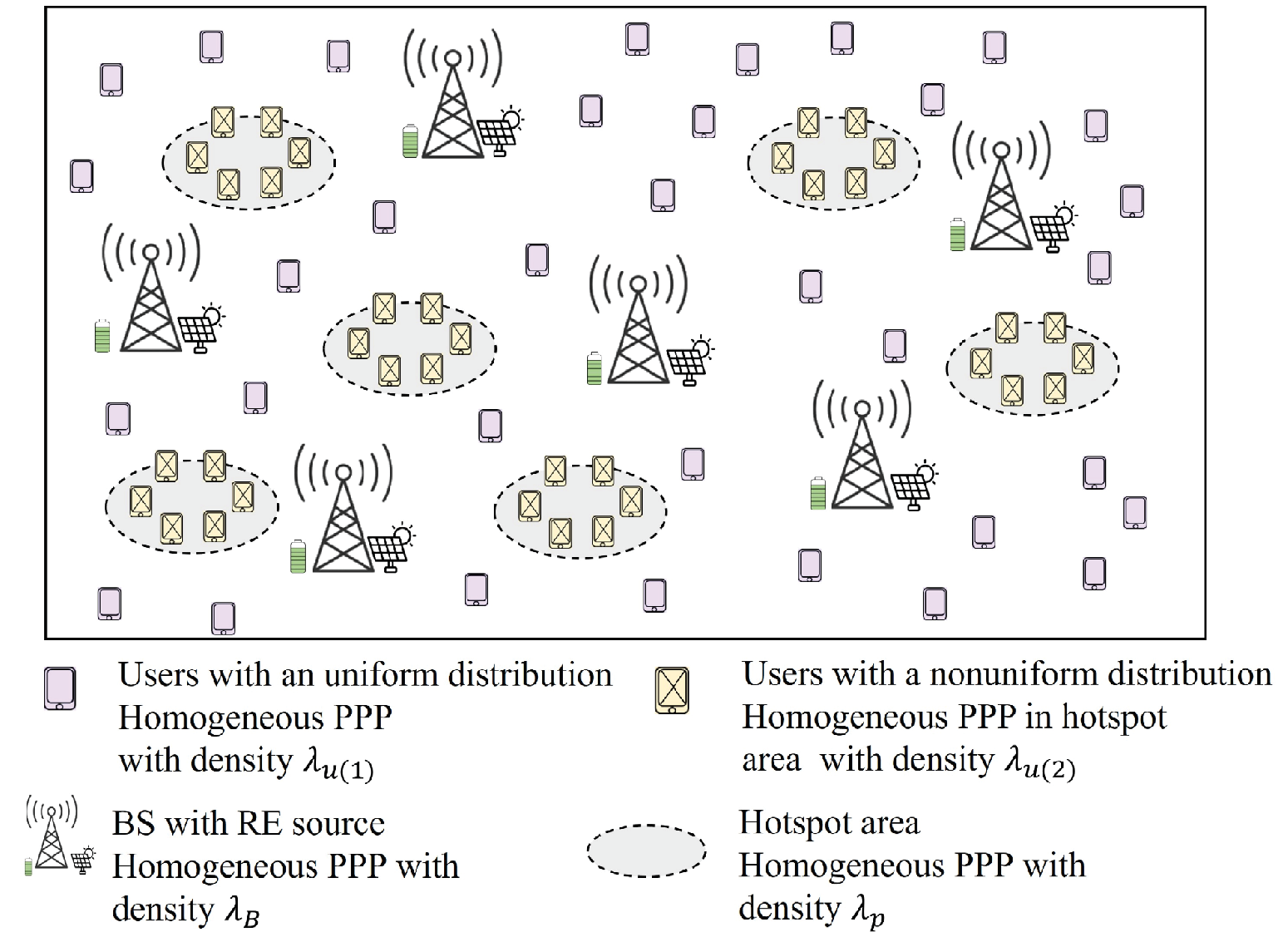}
\caption{System model.}
\label{fig_1}
\end{figure}

 Each BS is equipped with an RE generation system and a battery. The diagram of the network is shown in Fig. \ref{fig_1}.

Given that each BS is equipped with a battery, it is reasonable to assume that the energy generated by the distributed generation system of RE will be stored in the battery, and the consumption of RE by the BS will result in a reduction of the battery's charge, so we can use the RE state of the battery to model the RE state of the BS. The energy state of the battery is in discrete units, where one discrete unit of energy equals $\theta$ joules. Thus, the RE state of the battery is represented by $0, 1, \ldots, T$, which means that the amount of energy in the battery is $0, \theta, 2\theta, \ldots, T\theta$ joules. The RE state of the batteries allows for the classification of BSs into $T$ distinct types. We will use $\text{BS}^i, i=0,1,\ldots,T$ to denote a cell that is in RE state $i$.

In order to reduce the CEm of the network, a BS-management system is in place. This system is able to monitor the RE state and  capable of modifying the cell association bias parameters in the BSs, thereby enabling the adjustment of the number of users associated with the BS.

The energy consumption of the BSs can be divided into two categories: dynamic and static power consumption  \cite{auer_how_2011, israr2021renewable}. The energy consumption model is applicable to different types of BSs,  though the value of dynamic and static power consumption of each is distinct.
Dynamic power consumption depends on the load, while static power consumption is unaffected by the BS load. The power-consumption model of $\text{BS}^i$ is 

\begin{equation}
\label{eq:Pi}
P_{i}=P_0+\Delta_P P_t n_{i},
\end{equation}
where $P_0$ represents the static power consumption of the BS, $\Delta_P$ is the coefficient related to the power amplifier, $P_t=\frac{P_{\text{trans}}}{N}$ is the transmission power of one channel, $P_{\text{trans}}$ represents the transmission power of the BSs, $N$ is the total number of channels within a single BS, $n_i$ is defined as the number of the occupied channels in $\text{BS}^i$. We define $\theta \triangleq \Delta_P P_t \delta_t$, $\delta_t$ as the unit time.

\section{Modeling and analysis of RE-powered cellular networks}

In this section, we model the channel occupancy, the average number of users associated with a BS, the probability of successful transmission, and the user throughput.

\subsection{Channel-occupancy model}
A two-dimensional state is employed to describe the status of each BSs, wherein the first dimension denotes the RE state of the battery and the second represents the number of occupied channels. The user's service duration is assumed to be governed by exponential distributions with a mean $1/\mu$. The generation of energy units of RE is assumed to follow a Poisson process with rate $\nu$. The energy consumption by each occupied channel is represented by a Poisson process with rate $\omega$ \cite{dhillon_fundamentals_2014}. Moreover, since an energy-based cell association scheme is employed, the user-arrival rate depends on the RE state in the battery of the BS. The user-arrival process when the battery level is $i$ is modeled as a Poisson process with rate $\rho_i$, where $\rho_i$ depends on the mean number of users associated with $\text{BS}^i$, $U_i$, i.e., $\rho_i=h(U_i)$. Based on the above, the state of a BS is modeled as a QBD, whose transition-rate diagram is shown in Fig. 2.

This model can be extended to various scenarios, including mixed-energy networks, by adjusting the generation process of RE energy to different stochastic processes or by modifying the parameters of the Poisson process. For example, wind energy is often modeled as a different type of stochastic process  (e.g., in \cite{sturt2011efficient, verdejo2016stochastic}). This model also can be extended to heterogeneous network, different types of BSs have distinct energy generation and consumption parameters, and the state of these different types of BSs can be modeled accordingly.

\begin{figure}[!t]
\centering
\includegraphics[width=3.3in]{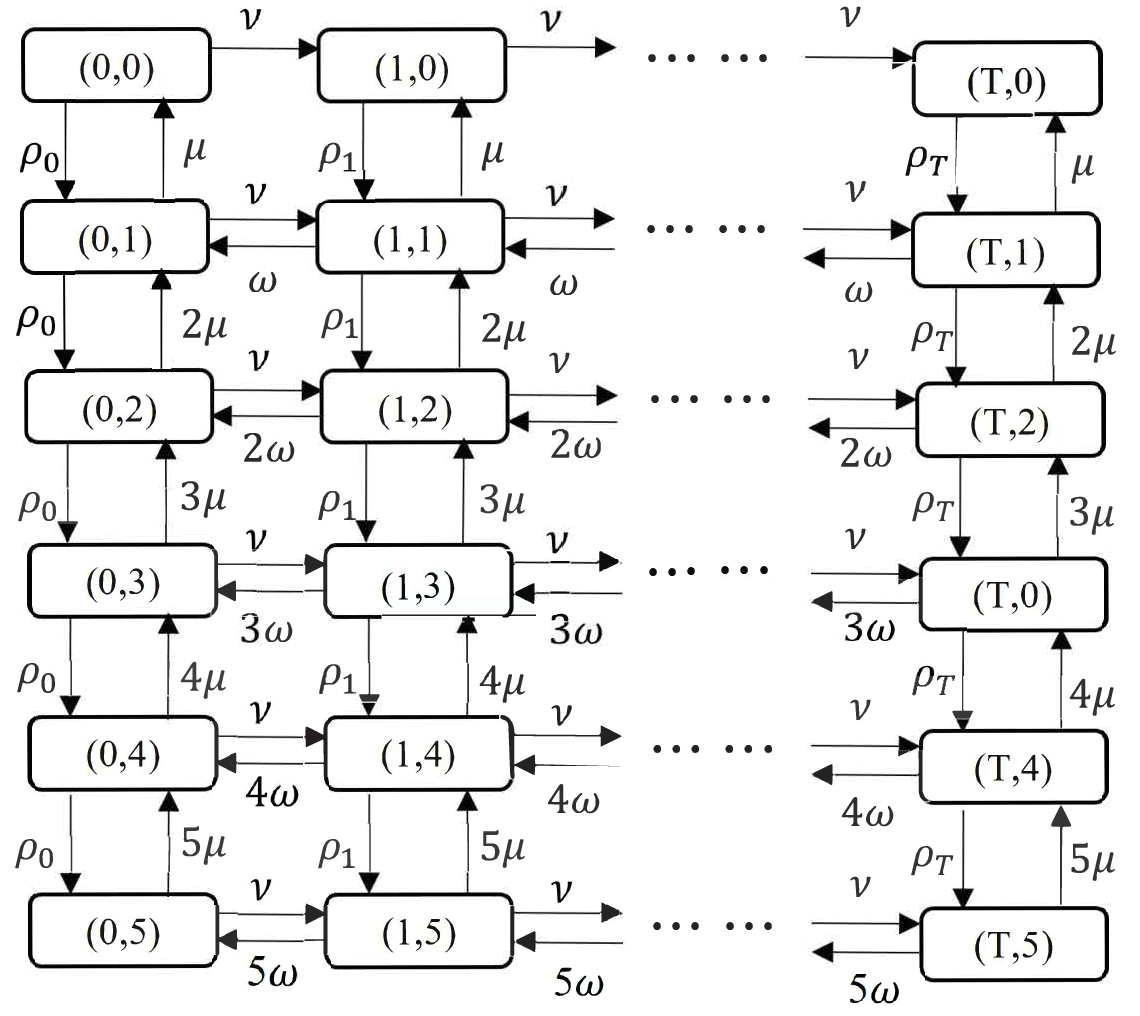}
\caption{State-transition diagram of the QBD with $N=5$ channels}
\label{fig2}
\end{figure}

Arranging the states in lexicographic order, the infinitesimal generator of the QBD $\mathbf{A}$, can be written as

\begin{equation}
\label{e2}
\mathbf{A}=\left[\begin{array}{cccccc}D_0 & L_0 & & & & \\ M_1 & D_1 & L_1 & & & \\ & M_2 & D_2 & L_2 & & \\ & & \ddots & \ddots & \ddots & \\ & & & M_{T-1} & D_{T-1} & L_{T-1} \\ & & & & M_T & D_T\end{array}\right],
\end{equation}
where $\mathbf{D}_i$, $\mathbf{M}_i$ and $\mathbf{L}_i$, 
for $i=0,1,2,\ldots,T$,
are square matrices of order $N+1$
that can be written as

\begin{equation}
\label{e3}
\mathbf{L}_i=\nu\mathbf{I}, i=0,1,\ldots,T-1,
\end{equation}
where $\mathbf{I}$ is the identity matrix with suitable dimensions,

\begin{equation}
    \label{e4}
    \mathbf{M}_i=\frac{P_0}{\theta}\mathbf{I}+\omega\,\mathrm{diag}(0,1,2,\ldots,N), 
     \ i=0,1,\ldots,T,
\end{equation}

\begin{equation}
\label{e5}
 \mathbf{D}_i=\left[\begin{array}{ccccc} *_{i,0}& \rho_i & & & \\ \mu &*_{i,i}&\rho_i &  & \\ &2 \mu &\ddots& \ddots & \\ & & \ddots&\ddots& \rho_i\\ & & & N \mu & *_{i,N} \end{array}\right], i=0,1,\ldots,T,
\end{equation}
where the $*_{i,j}$'s,
for $i=0,1,2,\ldots,T$ and $j=1,2, \ldots, N$, 
can be calculated by taking into account that the rows of the matrix $\mathbf{A}$ must sum up to 1 or, equivalently, from $(\mathbf{D}_0+\mathbf{L}_0)\mathbf{e}=0,(\mathbf{M}_i+\mathbf{D}_i+\mathbf{L}_i)\mathbf{e}=0,i=1,2,\ldots,T-1, (\mathbf{M}_T+\mathbf{D}_T)\mathbf{e}=0$, where $\mathbf{e}$ is a column vector of ones with appropriate dimensions.

We denote by $\pi(i,j)=\lim_{t\rightarrow \infty}\mathrm{P}_r[m(t)=i,n(t)=j]$ the probability that the BS is in state $(i,j)$, and $\boldsymbol{\pi}_i=\left(\pi(i,0),\pi(i,1),\ldots,\pi(i,N)\right),i = 0, 1, \ldots, T$. 
According to  the global balance equations, we obtain

\begin{equation}\label{e6}
    \begin{aligned}
        \boldsymbol{\pi}_0 \mathbf{D}_0+\boldsymbol{\pi}_1 \mathbf{M}_1 &= \mathbf{0} \\ 
        \boldsymbol{\pi}_0 \mathbf{L}_0+\boldsymbol{\pi}_1 \mathbf{D}_1+\boldsymbol{\pi}_2 \mathbf{M}_2 &=\mathbf{0} \\ 
        &~\vdots \\ 
        \boldsymbol{\pi}_{T-2} \mathbf{L}_{T-2}+\boldsymbol{\pi}_{T-1} \mathbf{D}_{T-1}+\pi_T \mathbf{M}_T &=\mathbf{0} \\ 
        \boldsymbol{\pi}_{T-1} \mathbf{L}_{T-1}+\boldsymbol{\pi}_T \mathbf{D}_T &=\mathbf{0},
    \end{aligned}
\end{equation}
where $\mathbf{0}$ is a row vector of size $N + 1$ with all its elements equal to zero.

By rearranging  (\ref{e6}), we obtain 
\begin{equation}\label{e7}
  \begin{aligned}
    \boldsymbol{\pi}_T &=-\boldsymbol{\pi}_{T-1} \mathbf{L}_{T-1} \mathbf{D}_T^{-1} \\
    \boldsymbol{\pi}_{T-1} &=-\boldsymbol{\pi}_{T-2} \mathbf{L}_{T-2} \mathbf{Q}_{T-1}^{-1} \\
    \boldsymbol{\pi}_{T-2} &=-\boldsymbol{\pi}_{T-3} \mathbf{L}_{T-2} \mathbf{Q}_{T-2}^{-1} \\
      &~\vdots \\
    \boldsymbol{\pi}_1&=-\boldsymbol{\pi}_0 \mathbf{L}_0 \mathbf{Q}_1^{-1} \\
    \boldsymbol{\pi}_0&\mathbf{Q}_{0}=0, 
  \end{aligned} 
\end{equation}  
where
\begin{equation}\label{eq:e7b}
  \begin{aligned}
    \mathbf{Q}_T&=\mathbf{D}_T \\
    \mathbf{Q}_{T-1}&=\mathbf{D}_{T-1}-\mathbf{L}_{T-1}\mathbf{Q}_{T}^{-1}\mathbf{M}_{T} \\
    \mathbf{Q}_{T-2}&=\mathbf{D}_{T-2}-\mathbf{L}_{T-2}\mathbf{Q}_{T-1}^{-1}\mathbf{M}_{T-1} \\
     &~\vdots \\
    \mathbf{Q}_{1}&=\mathbf{D}_{1}-\mathbf{L}_{1}\mathbf{Q}_{2}^{-1}\mathbf{M}_{2}  \\
    \mathbf{Q}_{0}&=\mathbf{D}_{0}-\mathbf{L}_{0}\mathbf{Q}_{1}^{-1}\mathbf{M}_{1}
  \end{aligned} 
\end{equation}



The last equation in~\eqref{e7}, $ \mathbf{Q}_{0}$ is an infinitrdimal generator, so it represents a linear system of equations
whose solution space has dimension~1; let $\hat{\boldsymbol{\pi}}_0$ be a solution to this system.
Then, $\hat{\boldsymbol{\pi}}_1, \hat{\boldsymbol{\pi}}_2, \ldots, \hat{\boldsymbol{\pi}}_T$
can be obtained by backward substitution in~\eqref{e7}.
Finally, the steady-state probability vectors of the QBD are given by
$\boldsymbol{\pi}_i = \hat{\boldsymbol{\pi}}_i/K$, for $i = 0, \ldots, T$,
where $K=\sum_{i=0}^{T}\hat{\boldsymbol{\pi}}_i$ is derived from the normalization condition.


Let $\pi_{i(\text{sum})}=\sum_{j=0}^{N}\pi(i,j)$.
It is noted that $\pi_{i(\text{sum})}$, for $i = 0, \ldots, T$, 
depend on the $\rho_k$'s ($k =0, \ldots, T$),
which in turn depend on the number of users associated with the BS, 
$\mathbf{U}=(U_0, U_1, \ldots, U_T)$.
Thus, we can write 
$\pi_{i(\text{sum})}=f_i(\mathbf{U})$, 
where 
$f_i: \mathbb{R}^{(T+1) \times 1} \rightarrow \mathbb{R}$ 
is the function that maps $U$ to $\pi_{i(\text{sum})}$.

The channel-blocking probability of $\text{BS}^i$ can be defined as the probability that all the channels are occupied, given that the energy level of the BS is $i$. This can be calculated as 
\begin{equation}
\label{e8}
    P_{\text{block},i}=\frac{\pi(i,N)}{\pi_{i(\text{sum})}}.
\end{equation}

The mean number of occupied channels of $\text{BS}^i$ is
\begin{equation}
\label{e10}
    n_i=\sum_{j=0}^Nj\frac{\pi(i,j)}{\pi_{i(\text{sum})}},
\end{equation}
and the probability that any given channel is occupied is given by
\begin{equation}
\label{e9}
    P_{\text{occu},i} =  \frac{n_i}{N}
\end{equation}

\subsection{The average number of associated users}
It is assumed that the state of each BS is independent from the state of the other BSs. Though  the states of BSs are not entirely independent and  this assumption may introduce some approximation errors, it provides an effective framework for theoretical analysis. Future work could explore incorporating the interdependencies between BSs to enhance the accuracy and practical applicability of the model in more complex network environments. So $\text{BS}^i, i=0,1,\ldots,T$ follow a homogeneous PPP, $\Phi_{B(i)}$, with the density $\lambda_{B(i)}=\pi_{i(\text{sum})}\lambda_B$. 
According to  Slivnyak's theorem \cite{haenggi_stochastic_2012}, 
for a user $u_0$ located at the origin of the coordinates,
the received power from $\text{BS}^i_z, z\in\Phi_{B(i)}$ is 
\begin{equation}
\label{e11}
P^i_{z,\text{receive}}=P_{t}h_z^i\left(d_z^i\right)^{-\alpha},
\end{equation}
where $h_z^i \sim \text{exp}(1)$ represents the Rayleigh fading of the channel between $\text{BS}^i_z$ and $u_0$, $d^i_z$ is the distance between $\text{BS}^i_z$ and $u_0$, and $\alpha$ is the path-loss coefficient.

In this paper, we propose a biased maximum long-term received-power cell association scheme. Let us denote as $\text{BS}^i_{z^{*}}$ the BS associated with $u_0$, that is

\begin{equation}\label{e12}
    \text{BS}^i_{z^{*}}=\underset{i=0,1, \ldots, T; z \in \Phi_{B(i)}}{\text{arg max}}B_i P_{t} \left(d_{z^*}^{i}\right)^{-a},
\end{equation}
where $B_i$ is the association bias of $\text{BS}^i$, which is a value related to the RE state of $\text{BS}^i$. Denoting $\mathbf{B}=\left(B_0,B_1,\ldots,B_T\right)\in \mathbb{R}^{(T+1)\times 1}$ for simplicity. 

The biased maximum long-term received-power cell association scheme maintains backward compatibility with conventional maximum long-term received-power association frameworks in practical deployment. Its operational workflow aligns with established cellular network architectures, requiring only software updates to existing base station controllers. The key enhancement lies in the integration of dynamic bias adjustment based on real-time RE state, where each BS periodically adjusts its association bias Bi. These optimized bias parameters are subsequently embedded into standardized cell selection signaling procedures, mirroring conventional system information block broadcast mechanisms defined in 3GPP TS 36.331\cite{3GPP_TS_38_104}

Based on Lemma 1 and Lemma 3 in \cite{jo_heterogeneous_2012},  the probability of $u_0$ being associated with $\text{BS}^i$ and the probability distribution function (PDF) of the distance between $u_0$ and $\text{BS}^i_{z^{*}}$ are, respectively,

\begin{equation}\label{e13}
    P_{\text{associa},i}=\frac{\lambda_{B(i)}B_i^{\frac{2}{\alpha}}}{\sum_{j=0}^T\lambda_{B(j)}B_j^{\frac{2}{\alpha}}},
\end{equation}


\begin{equation}\label{e14}
f_{d_{z^*}^{i}}(x)=\frac{2\pi\lambda_{B(i)}}{ P_{\text{associa},i}}x \text{exp}\left\{-\frac{\pi\lambda_{B(i)}}{P_{\text{associa},i}}x^2\right\}.
\end{equation}

The SINR of the channel between $\text{BS}_z^i$ and $u_0$ is

\begin{equation}\label{e15}
    \mathrm{SINR}_{\mathrm{BS}^i_{z^*}}(x)=\frac{P_t h^i_{z^*}x^{-\alpha}}{\sigma^2+\sum_{j=0}^T\underset{\mathrm{BS}_z^j\in\hat{\Phi}_{B(j)}\setminus \mathrm{BS}^i_{z^*}}{\sum}P_t h_z^i\left(d_z^i\right)^{-\alpha}},
\end{equation}
where $\sigma^2$ is the power of the additive white Gaussian noise, $\hat{\Phi}_{B(j)},j=0,1,\ldots,T$ is a homogeneous PPP with density $\lambda_{B(j)}P_{\mathrm{occu},j}$ that represents the set of BSs that has the interference on the same channel between $\mathrm{BS}_z^i$ and $u_0$.

The average number of users associated with $\mathrm{BS}^i$ is calculated as follows. For users with a nonuniform distribution, the PMF of the number of users in the circle area $S_{y_k}$ with radius $r$ at the center of the cluster $y_k \in \Phi_p$ is

\begin{equation}\label{e16}
    \mathrm{P}_r[\Phi_{u(2),k}(S_{y_k})=a]=\frac{\left(\lambda_{u(2)}\pi r^2\right)^a}{a!}e^{-\lambda_{u(2)}\pi r^2}.
\end{equation}

It is assumed that all users within a cluster associate with the same BS. And we denote $S_i$ as the coverage area of a BS that belongs to $\mathrm{BS}^i$.
Thus, the PMF of the number of uniformly distributed users with in  $S_i$ is 
\begin{multline}\label{e17}
    \mathrm{P}_r[\Phi_{u(2)}(S_{i})=a]\\
      =\sum_{b=0}^{\infty}\mathrm{P}_r[\Phi_{p}(S_{i})=b]\left[\mathrm{P}_r[\Phi_{u(2),k}(S_{i})=a|\Phi_{p}(S_{i})=b]\right]\\
      =\sum_{b=0}^{\infty}\frac{\left(\lambda_pS_{i}\right)^b}{b!}e^{-\lambda_pS_{i}}\left(\frac{\left(b\lambda_{u(2)}\pi r^2\right)^a}{a!}e^{-b\lambda_{u(2)}\pi r^2}\right).
\end{multline}

The average number of users in the area $S_{i}$ is given as
\begin{multline}\label{e18} 
 \sum_{a=0}^{\infty}a\mathrm{P}_r[\Phi_{u(2)}(S_{i})=a]\\
    =\sum_{b=0}^{\infty}\left(\frac{e^{-\lambda_pS_{i}}}{b!}\left(\lambda_p S_{i}\right)^b
     \left(\sum_{a=0}^{\infty}a\frac{\left(b\lambda_{u(2)}\pi r^2\right)^{a}}{a!}
     e^{-b\lambda_{u(2)}\pi r^2}\right)\right)\\
    =\sum_{b=0}^{\infty}\left(\frac{e^{-\lambda_pS_{i}}}{b!}\left(\lambda_p S_{i}\right)^bb
      \lambda_{u(2)}\pi r^2\right)
    =\lambda_{u(2)}\pi r^2\lambda_p S_{i},
\end{multline}

Now, since $\mathbf{E}\left[S_{i}\right]=P_{\mathrm{associa},i}/\lambda_{B(i)}$ \cite{dhillon_fundamentals_2014}, we can write

\begin{equation}\label{e19}
        U_{i(2)}=\lambda_{u(2)}\pi r^2\lambda_p \mathbf{E}[S_{i}]=\frac{\lambda_p\lambda_{u(2)}\pi r^2 B_i^{\frac{2}{\alpha}}}{\sum_{j=0}^T\lambda_{B(j)}B_j^{\frac{2}{\alpha}}}.
\end{equation}

For uniformly distributed users, the average number of users in the area $S_{i}$ is 
\begin{multline}\label{e20}
    U_{i(1)} = \\
     \mathbf{E}[S_{i}]\lambda_{u(1)}=\frac{P_{\mathrm{associa}}\lambda_{u(1)}}{\lambda_{B(i)}}=\frac{\lambda_{u(1)}B_i^{\frac{2}{\alpha}}}{\sum_{j=0}^T\lambda_{B(j)}B_j^{\frac{2}{\alpha}}}.
\end{multline}

The average number of users in the area $S_{i}$ is
\begin{multline}\label{e21}
    U_i =U_{i(1)}+U_{i(2)}=\frac{\left(\lambda_{u(2)}\pi r^2 \lambda_p+\lambda_{u(1)}\right) B_i^{\frac{2}{\alpha}}}{\sum_{j=0}^T \lambda_{B(j)} B_j^{\frac{2}{\alpha}}} \\
     =\frac{\left(\lambda_{u(2)}\pi r^2 \lambda_p+\lambda_{u(1)}\right) B_i^{\frac{2}{\alpha}}}{\lambda_B \sum_{j=0}^T \pi_{j(\mathrm{sum})} B_j^{\frac{2}{\alpha}}} 
    = g_i(\mathbf{\Pi}),
\end{multline}
where 
$\mathbf{\Pi} = 
  \left(\pi_{0(\mathrm{sum})}~\pi_{1(\mathrm{sum})}~\cdots~\pi_{T(\mathrm{sum})}\right)^\top 
    \in \mathbb{R}^{(T+1)\times 1}$ 
and 
$g_i:\mathbb{R}^{(T+1)\times 1}\rightarrow\mathbb{R}$ is a function from $U_i$ to $\mathbf{\Pi}$.

Now, we introduce 
$F(\mathbf{U}) = \left(f_0(\mathbf{U})~\cdots~f_T(\mathbf{U})\right)^\top$
and
$G(\mathbf{\Pi}) = \left(g_0(\mathbf{\Pi})~\cdots~g_T(\mathbf{\Pi})\right)^\top$
for notational convenience.
Thus, 
\begin{align}
    \mathbf{\Pi} &= F(\mathbf{U}) \label{eq: Pi_F_U} \\
    \mathbf{U} &= G(\mathbf{\Pi}) \label{eq: U_G_Pi},
\end{align}
which capture the coupling relationship between $\mathbf{U}$ and $\mathbf{\Pi}$.
Clearly, if there exist $\mathbf{U}$ and $\mathbf{\Pi}$ satisfying these two equations simultaneously,
they are fixed points of $\Xi = G \circ F$ and $\Gamma = F \circ G$, respectively:
i.e., $\mathbf{U} = \Xi(\mathbf{U})$ and $\mathbf{\Pi} = \Gamma(\mathbf{\Pi})$.



If the function $\Gamma$ is a continuous mapping from a nonempty, closed, bounded and convex set to itself,
it follows that the solution $\mathbf{\Pi}$ must exist,
according to Brouwer's fixed-point theorem~\cite{milnor_analytic_1978}.
It is not possible to obtain an explicit form of the function $\Gamma$. 
Nevertheless, $\Gamma$ defines a mapping from $[0, 1]^{(T+1)\times 1}$ to itself,
and it seems intuitive ---and consistent with our numerical experiments---
that it is continuous. 
Therefore, if we admit the assumption that $\Gamma$ is continuous,
we can state the existence of the fixed point $\mathbf{\Pi}$.
We calculate a numerical approximation of the fixed points $\mathbf{\Pi}$ and $\mathbf{U}$
through the iterative procedure outlined in Algorithm~\ref{alg:alg1}, 
which has proven to converge in all of our numerical experiments.

\begin{algorithm}[tbh]
\caption{Iterative algorithm to calculate $\mathbf{\Pi}$ and $\mathbf{U}$}\label{alg:alg1}
\begin{algorithmic}
\STATE 
\STATE \textbf{Input:}   
  RE generation rate, $\nu$; 
  user-service rate, $\mu$;  
  energy-consumption rate, $\omega$; 
  number of channels, $N$;  
  density of BSs, $\lambda_B$;
  density of uniformly dist. users, $\lambda_{u(1)}$;
  density of social hotspots, $\lambda_{p}$;
  density of nonuniformly dist. users; $\lambda_{u(2)}$;
  radius of social hotspots, $r$;
  path-loss coefficient, $\alpha$;
  average number of users, $\mathbf{U}$;
  steady-state probabilities of BSs, $\mathbf{\Pi}$;
  association bias vecto, $\mathbf{B}$;
  maximum number of iterations, $t_{\max}$;
\STATE \textbf{Output:} 
  Average number of users, $\mathbf{U}$; 
  steady-state probabilities of BSs $\mathbf{\Pi}$;
\STATE \hspace{0.5cm}\textbf{Initialization:} $\mathbf{\Pi}^{(0)}$ 
\STATE \hspace{0.5cm} $t=0$,  $\mathbf{U} = G(\mathbf{\Pi}^{(0)})$
\STATE \hspace{0.5cm} \textbf{Repeat}
\STATE \hspace{1cm}     $t=t+1$
\STATE \hspace{1cm}     $\mathbf{\Pi} = F(\mathbf{U})$ 
\STATE \hspace{1cm}     $\mathbf{U} = G(\mathbf{\Pi})$ 
\STATE \hspace{1cm}     $d = \left\| \mathbf{\Pi} - \mathbf{\Pi}^{(0)}\right\|$
\STATE \hspace{1cm}     $\mathbf{\Pi}^{(0)} = \mathbf{\Pi}$ 
\STATE \hspace{0.5cm}\textbf{Until:} $d<\varepsilon$ or $t=t_{\max}$
\end{algorithmic}
\end{algorithm}

\subsection{Probability of successful transmission}
The probability of a successful transmission is defined as the probability that the SINR of the channel is greater than the received  threshold $\tau$, which is calculated as
\begin{multline}
\mathrm{P}_{\text {succ}, i}(\tau)
=\int_{x=0}^{\infty} \mathrm{P}_{\mathrm{r}}\left[\mathrm{SINR}_{i}(x)>\tau\right] f_{d_i^i}(x)x d x\\
=\frac{2 \pi \lambda_{B(i)}}{P_{\mathrm{associa},i}} \int_{x=0}^{\infty} x \mathrm{P}_{\mathrm{r}}\left[\mathrm{SINR}_{i}(x)>\tau\right]\\
\times  \exp\left\{-\frac{\pi\lambda_{B(i)}}{P_{\mathrm{associa},i}}x^2\right\} d x \\
 \stackrel{(a)}{=} \frac{2 \pi \lambda_{B(i)}}{P_{\mathrm{associa},i}} \int_0^{\infty} x \exp \left\{-\frac{\tau \sigma^2}{P_t x^{-\alpha}}-\pi C_i x^2\right\} d x,
 \label{e24}
\end{multline}
where
\begin{equation}
C_i=\sum_{j=1}^T \lambda_{B(j)}\left[\left(\frac{B_j}{B_i}\right)^{\frac{2}{\alpha}}+P_{\mathrm{occu}, j}\mathcal{Z}_i\left(\tau, \alpha, B_j\right)\right], \tag{\ref{e24}{a}} \label{e24a}
\end{equation}
and it has been used that (see \cite{jo_heterogeneous_2012} for a detailed derivation)
\begin{multline}
     \mathrm{P}_{\mathrm{r}}\left[\mathrm{SINR}_{i}(x)>\tau\right] = \\
     \exp \left(-\frac{\tau \sigma^2}{P_t x^{-\alpha}} - \pi \sum_{j=1}^T P_{\mathrm{occu},j}\lambda_{B(j)} 
    \mathcal{Z}_i\left(\tau,\alpha,B_j\right) x^2 \right),  
    \tag{\ref{e24}{b}} \label{e24b}
\end{multline}
%
with
\begin{multline}
\mathcal{Z}_i\left(\tau,\alpha,B_j\right)\\
=\frac{2\tau\left(\frac{B_j}{B_i}\right)^{\frac{2}{\alpha}-1}}{\alpha-2}_2F_1\left[1,1-\frac{2}{\alpha};2-\frac{2}{\alpha};-\frac{\tau B_j}{B_i}\right],
\tag{\ref{e24}{c}} \label{e24c}
\end{multline}
and ${ }_2 F_1[\bullet,\bullet;\bullet;z]$ is a Gaussian hypergeometric function defined for $|z|< 1$. 
In light of the fact that interference occurs exclusively between BSs utilizing the same channel, the density of BSs causing interference in this paper is $\lambda_{B(j)}P_{\mathrm{occu},j}$.

The successful transmission probability of the typical user $u_0$ is

\begin{equation}\label{e25}
    \mathrm{P}_{\mathrm{succ}}(\tau)=\sum_{i=0}^{T}\mathrm{P}_{\mathrm{succ},i}(\tau)P_{\mathrm{associa},i},
\end{equation}
where $\mathrm{P}_{\mathrm{succ},i}$ and $P_{\mathrm{associa},i}$ are defined in (\ref{e13}) and (\ref{e24})

\subsection{Throughput model}
The expected throughput of a typical user $u_0$ associated with $\mathrm{BS}^i$ is given by (\ref{e26}), displayed at the top of  the next page \cite{ge_spatial_2015}.
\begin{figure*}
\begin{equation}\label{e26}
\begin{aligned}
    \mathcal{R}_{i}(\tau)=&\left(1-P_{\mathrm{block},i}\right)\mathrm{P}_{\mathrm{succ},i}(\tau)\mathbf{E}\left[\log_2\left(1+\mathrm{SINR}_{i}\right)\right]\\
    =&\left(1-P_{\mathrm{block},i}\right)\mathrm{P}_{\mathrm{succ},i}(\tau)\int_0^{\infty}\int_0^{\infty}\mathbf{P}_r\left[\log_2\left(1+\mathrm{SINR}_{i}(x)\right)>t\right]dtf_{d_{z^*}^{i}}(x)dx\\
    =&\left(1-P_{\mathrm{block},i}\right)\mathrm{P}_{\mathrm{succ},i}(\tau)\int_0^{\infty}\int_0^{\infty}\mathbf{P}_r\left[\mathrm{SINR}_{i}(x)>2^t-1\right]dtf_{d_{z^*}^{i}}(x)dx\\
    =&\left(1-P_{\mathrm{block},i}\right)\mathrm{P}_{\mathrm{succ},i}(\tau)\int_0^{\infty}\mathrm{P}_{\mathrm{succ},i}\left(2^t-1\right)dt.
\end{aligned}
\end{equation}
\end{figure*}

Therefore, the total throughput per unit area of the network can be calculated as

\begin{equation}\label{e27}
    \mathcal{R}(\tau)=\left(\lambda_{u(2)}\pi r^2 \lambda_p+\lambda_{u(1)}\right)\sum_{i=0}^T\frac{\rho_i}{U_i}\mathcal{R}_{i}(\tau)P_{\mathrm{associa}, i},
\end{equation}
where $\frac{\rho_i}{U_i}$ is the average ratio of active users in $\mathrm{BS}^i$.

\subsection{Carbon efficiency of RE-powered cellular networks}
Based on the power-consumption model of $\mathrm{BS}^i$ shown in~\eqref{eq:Pi}, 
the average power consumption per unit area of the network is 

\begin{equation}\label{e28}
    P_{\mathrm{tot}}=\sum_{i=0}^T P_{i}\lambda_{B(i)}=\lambda_{B}\sum_{i=0}^T P_{i}\pi_{i(\mathrm{sum})}.
\end{equation}

The energy efficiency, expressed in $\text{bps}/(\text{J}\times\text{unit area}^{-1})$  can be modeled as 
\begin{equation}\label{e29}
   \eta_{\mathrm{ee}}=\frac{\mathcal{R}(\tau)\delta_t}{P_{\mathrm{tot}}\delta_t}=\frac{\mathcal{R}(\tau)}{P_{\mathrm{tot}}}.
\end{equation}
The CEm factors \cite{eggleston_2006_2006} defined by the Intergovernmental Panel on Climate Change (IPCC) are used to calculate the CEm of the BSs. In the network, the use of the grid source is contingent upon the battery state of the BSs: a BS will only demand energy from the grid when its battery is in state 0. Consequently, the grid power consumed per unit area of the network is 

\begin{equation}\label{e30}
    P_{\mathrm{grid}}=P_{0}\lambda_{B(0)}=\left(P_0+\Delta_P P_t n_0\right)\lambda_{B(0)}.
\end{equation}

The CEm per unit area of the network in unit time, expressed in $\text{gCO}_2/\text{unit area}$ is then given as

\begin{equation}   \label{e31}
E_{\mathrm{tot}}=(P_{\mathrm{grid}}\xi_{\mathrm{grid}}+\sum_{i=1}^TP_{i}\lambda_{B(i)}\xi_{\mathrm{re}})\delta_t,
\end{equation}
where $\xi_{\mathrm{grid}}$ and $\xi_{\mathrm{re}}$ are the CEm factors of the grid source and RE source, respectively. 
Only the CE in the usage phase is considered in this paper, so the carbon-emission factor of the RE source is 0. In some cases, the life-cycle carbon emissions must be considered, where the carbon-emission factor of the RE source is greater than 0.

According to \cite{zhao_carbon_2024}, the CEf is defined as the ratio between the network throughput and the CEm, expressed in $\text{bps}/\text{gCO}_2$. The CEf is calculated as

\begin{equation}\label{e32}
        \eta_{\mathrm{ce}}=\frac{\mathcal{R}(\tau)\delta_t}{E_{\mathrm{tot}}}=\frac{\mathcal{R}(\tau)}{P_{\mathrm{grid}}\xi_{\mathrm{grid}}+\sum_{i=1}^T P_{i}\lambda_{B(i)}\xi_{\mathrm{re}}}.
\end{equation}
\section{Network-performance analysis}
According to the definition of the IPCC, the CEm factor of the grid source is considerably higher than that of the RE source. Therefore, to reduce the network’s CEm, it is essential to decrease the energy consumption of the grid source. In this section, we examine a simple design for the association bias parameters, assuming a functional form that depends on a single easily optimizable parameter. In the following section, we extend the discussion to a more complex general optimization framework. Let $B_i=u(i)$, where $u: \mathbb{Z}\rightarrow \mathbb{R}$ is a function satisfying the following two conditions:
{\begin{enumerate}
\item{$u(0)=1$;} 
\item{$u(i+1)>u(i) \quad i=0,1,2,\ldots,T-1$.}
\end{enumerate}}

Here, we propose  $B_i=u(i)=(i+1)^\beta$, where $\beta \in \mathbb{R}_{\geq 0}$ is a parameter that regulates the influence of the BS battery level on the association bias. When $\beta=0$, we have that $B_i=1$ for all i’s, and the cell association scheme is simplified to the nearest cell association scheme. Conversely, when $\beta>0$  the scheme becomes an energy-based cell association scheme. As $\beta$ increases, the bias also increases, implying that the influence of the battery level of the BSs on the cell association scheme grows with the value of $\beta$.

The function relating the user-arrival rate to the number of users associated with $\mathrm{BS}^i$, defined as $\rho_i=h(U_i)=U_i$, ensures that the order of magnitude of the user-arrival rate is consistent with that in \cite{massey2002analysis}. The values of the remaining parameters are provided in Table~\ref{tab2}.

\begin{table}[tbh]
\caption{Simulation Parameters}\label{tab2}
\renewcommand{\arraystretch}{1.2}
\begin{center}
\begin{tabular}{@{}lcp{0.2\columnwidth}@{}}
\toprule
 \textbf{Description} & \textbf{Symbol} & \textbf{Value} \\
\midrule
Static power consumption of BSs & $P_{0}$ & 56\,W\cite{auer_how_2011} \\
Coefficient related to the power amplifier& $\Delta_{P}$ & $2.6$ \cite{auer_how_2011} \\
Transmission power of BSs& $P_{\text{trans}}$ & $6.3$\,W \cite{auer_how_2011} \\
Density of BSs& $\lambda_B$ & $1\,\text{km}^{-2}$ \\
Density of uniformly distributed users& $\lambda_{u(1)}$ & $5\,\text{km}^{-2}$ \\
Density of social hotspots& $\lambda_p$ & $1\,\text{km}^{-2}$ \\
Density of nonuniformly dist. users& $\lambda_{u(2)}$ & $1\,\text{km}^{-2}$ \\
Radius of social hotspots area& $r$ & $2\,\text{km}$ \\
Capacity of BS battery& $T$ & 10 \\
Path-loss coefficient& $\alpha$ & 4 \cite{andrews_tractable_2011} \\
Additive white Gaussian noise& $\sigma^2$ & $-40\,\text{dBm} $ \\
Receiver threshold& $\tau$ &  $-10\,\text{dB} $ \\
User-service rate& $\mu$ &  $2\,\text{min}^{-1} $ \\
Energy-consumption rate& $\omega$ & 1 \\
Carbon-emission factor of grid source& $\xi_{\text{grid}}$ & $1.5842\times 10^{-4}$ \newline $\text{g}\text{CO}_2/\text{J}$ \cite{mee2023notice} \\
Carbon-emission factor of RE source& $\xi_{\text{re}}$ & $0\,\text{g} \text{CO}_2/\text{J}$ \\
Total number of channels& $N$ & 20 \cite{ge_spatial_2015} \\
\bottomrule
\end{tabular}
\end{center}
\end{table}

\subsection{Probability of successful transmission}

\begin{figure}[tb]
\centering
\includegraphics[width=3.5in]{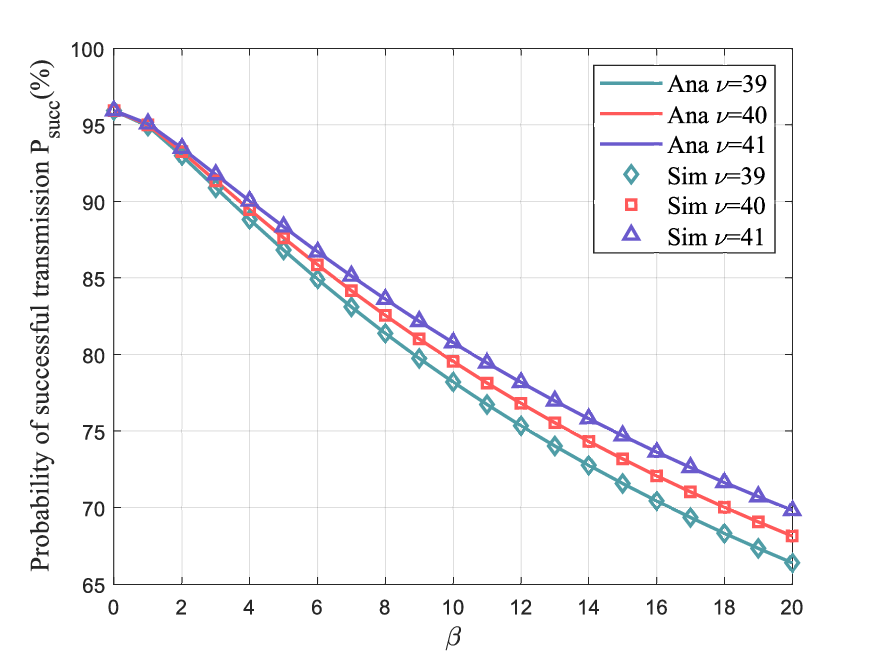}
\caption{Probability of successful transmission.}
\label{fig_3}
\end{figure}

\begin{figure*}[tb]
\centering
\subfloat[]{\includegraphics[width=3in]{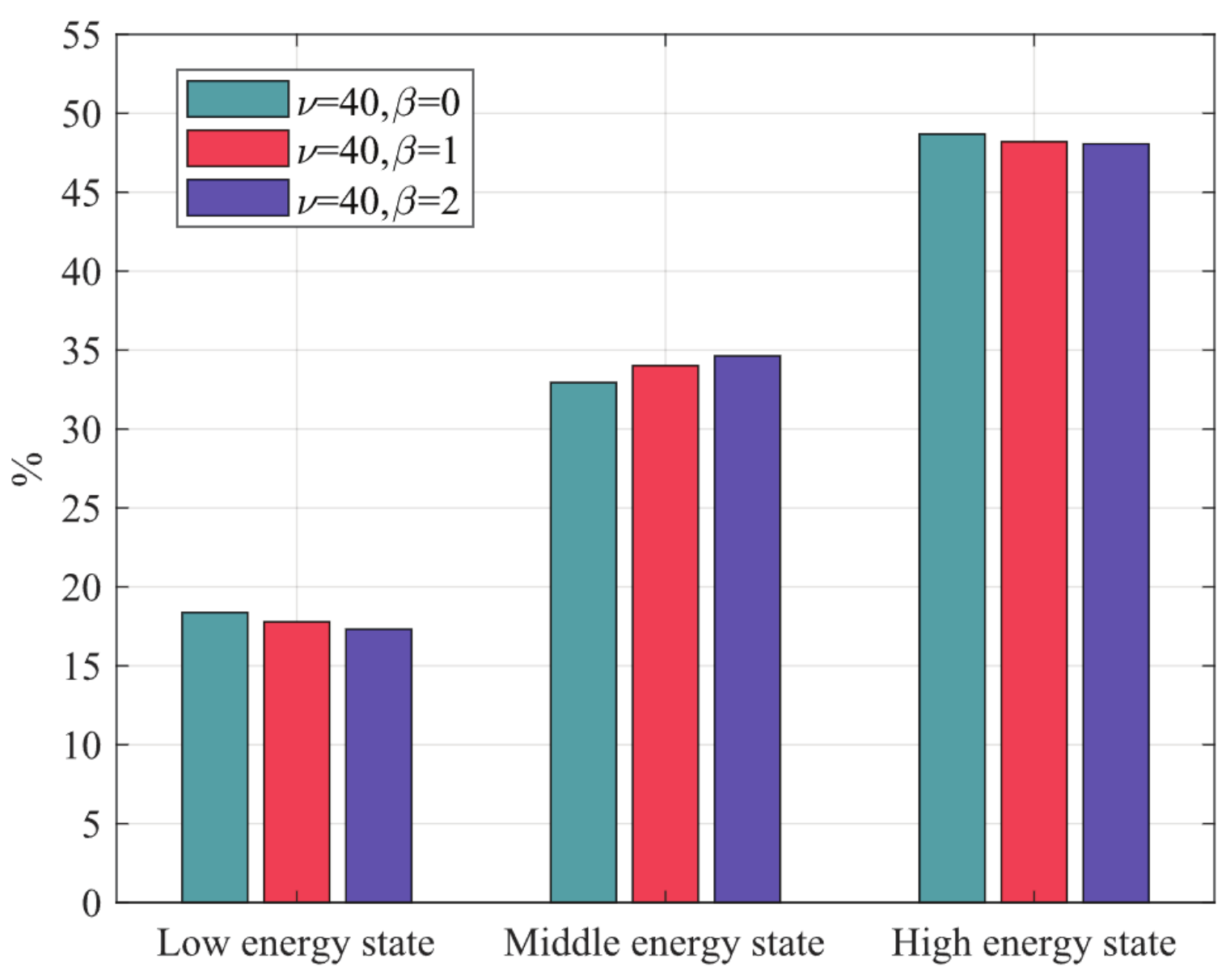}
\label{fig_4(a)}}
\hfil
\subfloat[]{\includegraphics[width=3in]{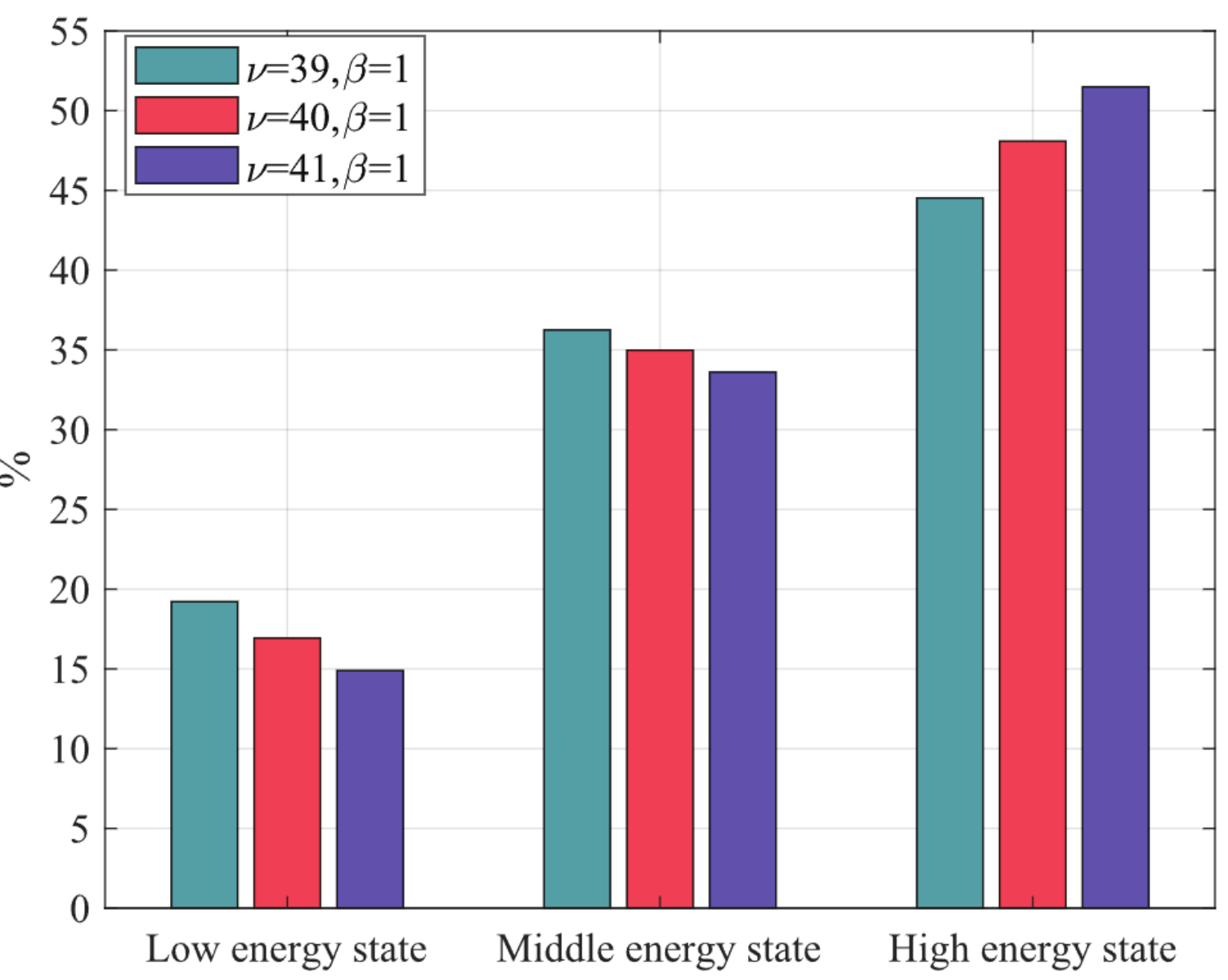}
\label{fig_4(b)}}
\caption{
  (a) Proportion of BSs in different statuses within different association-bias parameters. 
  (b) Proportion of BSs in different statuses within different RE-generation rates.
}
\end{figure*}

The results of (\ref{e25}) are compared with those obtained through a Monte-Carlo simulation to verify the reliability of the derivation. As shown in Fig. \ref{fig_3}, the analytical results are in good agreement with the Monte-Carlo simulation results, thereby verifying the correctness of the theoretical analysis presented in this paper.

In addition, when the RE-generation rate $\nu$ is different, the relationship  between the probability of successful transmission $\mathrm{P}_{\text{succ}}$ and the association-bias parameter $\beta$ is also depicted in Fig. \ref{fig_3}. When the RE-generation rate is maintained at a constant level, with the increase of the association-bias parameter $\beta$, the probability of successful transmission is reduced. Moreover, when the association-bias parameter remains constant, an increase in the RE-generation rate results in an elevated probability of successful transmission. This is due to the fact that the larger the association-bias parameter, the greater the dependence of the bias of cell association on the RE state of the BSs. As a result, users are more likely to associate with BSs having sufficient RE, rather than those with higher received power. This shift ultimately decreases the SINR, thereby decreasing the probability of successful transmission.

\subsection{Distribution of the battery level of BSs}

Fig. 4 shows the proportion of BSs with low (i = 0, 1, 2), medium (i = 3, 4, 5, 6), and high (i = 7, 8, 9, 10) battery levels for various values of the RE generation rate $\nu$ and the association-bias parameter $\beta$. 

Fig. 4(a) shows the proportion of BSs in the different battery level categories for various values of the association-bias parameter, with $\nu=40$. As shown,  a larger association-bias parameter results in a greater proportion of the BSs exhibiting a middle-RE state and a reduced proportion of the BSs displaying a low-RE state and a high-RE state. When $\beta=2$,  the proportion of BSs in the middle-RE state is greater than $\beta=1$ and $\beta=0$, and  when $\beta=0$,  the proportion of BSs in a low-RE state is greater than $\beta=1$ and $\beta=2$. This is due to the fact that as the association-bias parameter increases, users are more inclined to access BSs with a higher-RE state. This results in an increase in the power consumption of this part of the BSs, thereby causing the rapid utilization of energy. Conversely, BSs with a low-RE state demonstrate a reduction in power consumption due to a smaller number of associated users. Furthermore, the RE state of this part of the BSs increases because of the generation of RE. Consequently, the proportion of middle-energy-state BSs increases.

Fig. 4(b) shows the proportion of BSs in the different battery level categories for various values of the RE generation rate, with $\beta=1$. As the RE generation rate rises, the proportion of BSs with a high RE state also rises. This implies that as the RE generation rate rises, the number of BSs consuming the grid source decreases. This is because a higher RE generation rate makes it easier for BSs to reach a high-RE state. As expected, it is concluded that the RE generation rate affects the CEm of the network, since an increase in the RE generation rate results in a reduction in the CEm of the network.

\subsection{CEm and CEf}
\begin{figure}[tb]
\centering
\includegraphics[width=3in]{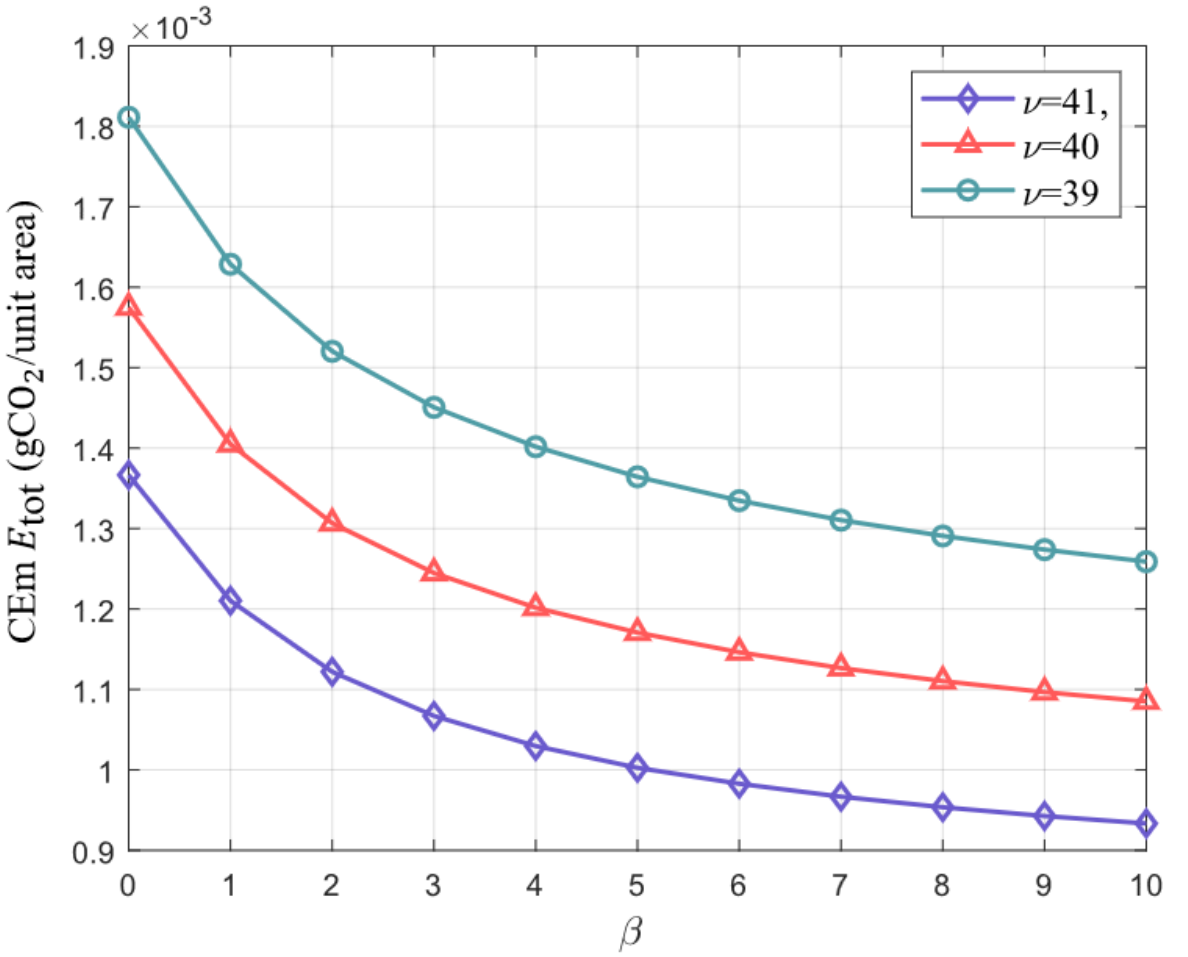}%
\label{fig_5}
\caption{ CEm per unit area.}
\end{figure}

The relationship between the CEm per unit area $E_{\text{tot}}$, and the association-bias parameter $\beta$ for a varying RE-generation rate $\nu$ is illustrated in Fig. 5. When the RE-generation rate is maintained at a constant level, an increase in the association-bias parameter results in a reduction of the CEm. When the association-bias parameter is held constant, an increase in the RE-generation rate results in a corresponding decrease in the CEm. This is due to the fact that as the association-bias parameter increases, users are more likely to associate with the BSs having sufficient RE. Consequently, the probability of the BSs with insufficient RE using the grid source decreases, which ultimately results in a reduction in the CEm of the network.

\begin{figure*}[tb]
\centering
\subfloat[]{\includegraphics[width=3in]{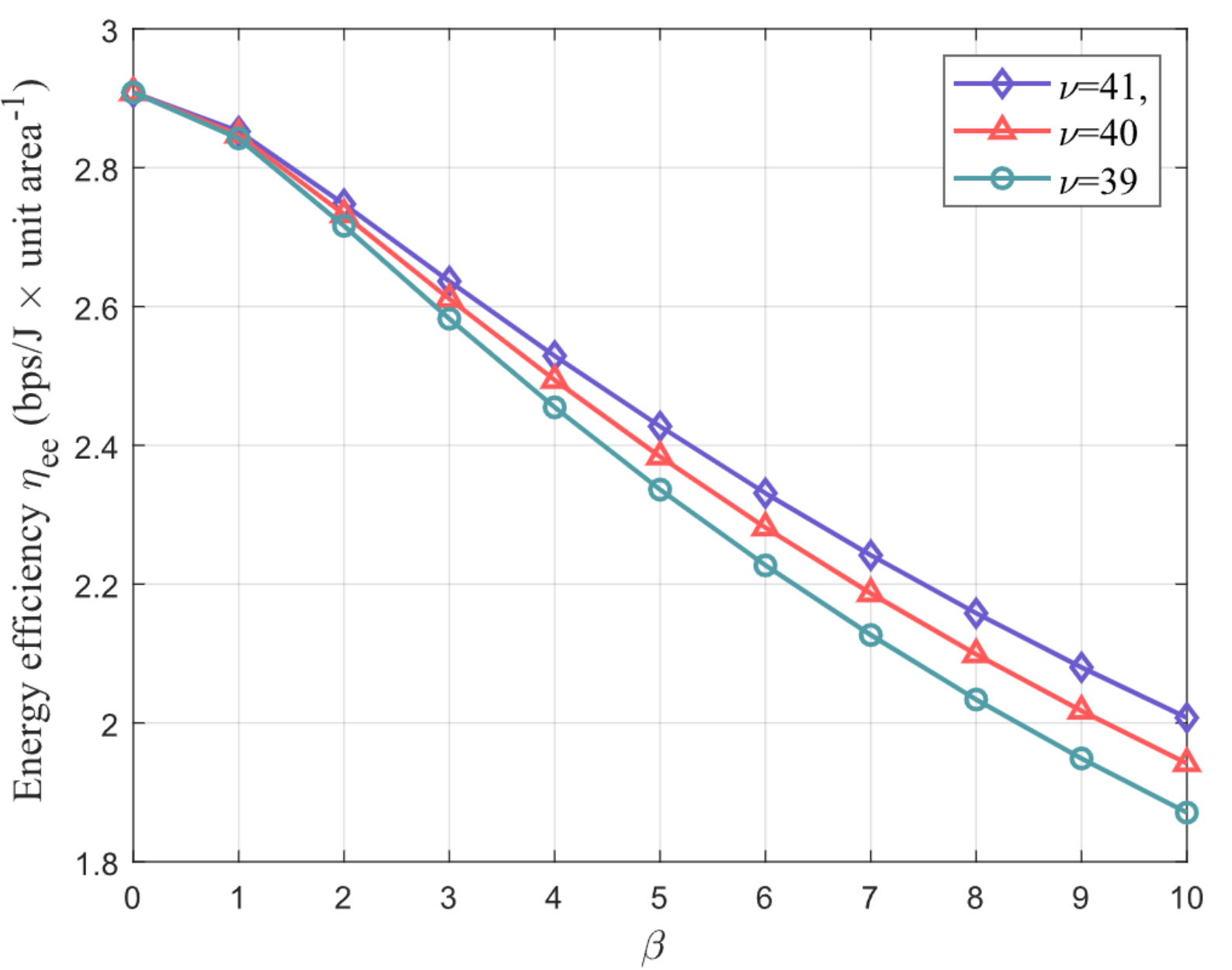}
\label{fig_6(a)}}
\hfil
\subfloat[]{\includegraphics[width=3in]{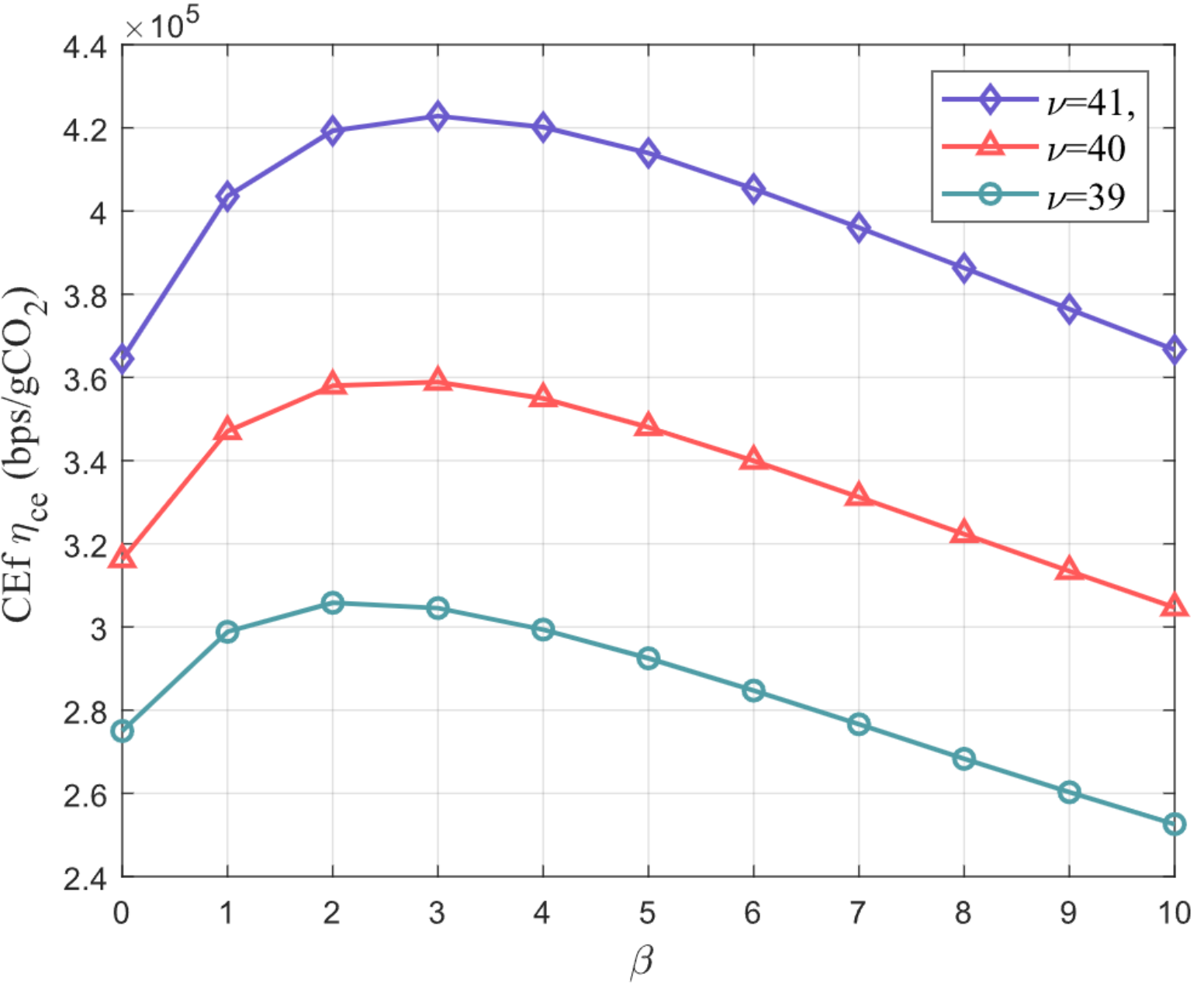}
\label{fig_6(b)}}
\caption{ Energy efficiency (a) and CEf (b) as functions of the parameter $\beta$ in the association bias for different values of the RE-generation rates ($\nu$).}
\end{figure*}

Fig. 6 shows the correlation between the energy efficiency $\eta_{\text{ee}}$, the CEf $\eta_{\text{ce}}$, and the association-bias parameter $\beta$. Fig. 6(a) shows that when the RE-generation rate is maintained at a constant level, an increase in the association bias results in a rapid decline in the energy efficiency of the network.  Conversely, when the association-bias parameter is maintained constant, an increase in the RE-generation rate results in an increase in energy efficiency. This is due to the fact that the larger the association-bias parameter, the lower the probability of successful transmission, which in turn leads to a reduction in the network throughput and energy efficiency. 

As shown in Fig. 6(b), when the RE-generation rate is maintained at a constant level, an increase in the association-bias parameter initially enhances the CEf, followed by a subsequent decline. There is an optimum value for the association-bias parameter that maximizes the CEf of the network. This is because, as the association-bias parameter increases, users are more inclined to associate with BSs with higher energy states. As a result, BSs have lower probability of using grid energy, leading to a decrease in the CEm and an increase in the CEf. However, an excessively high association-bias parameter can cause the cell association scheme to become overly dependent on the RE state of the BSs, neglecting the SINR. This may reduce the network's throughput and, consequently, the CEf. When the RE-generation rate takes the values $\nu=39$, 40 and 41, the optimum association-bias parameter is, respectively, $\beta=2$, 3 and 3, respectively. Furthermore, when the association-bias parameter is held constant, an increase in the RE-generation rate leads to an enhancement in the maximum CEf of the network, because the higher RE-generation rate can reduce the CEm of the network. Furthermore, compared to Fig. 6(a), it is evident that a decrease in the network's energy efficiency does not necessarily result in a lower CEf. This underscores the need to implement the CEf as a more relevant measure than energy efficiency.

The aforementioned results demonstrate that the CEm of the network can be diminished through the implementation of an energy-based cell association scheme. However, this approach inevitably entails a compromise in QoS. An optimal energy-based cell association scheme can be identified that simultaneously maximizes the CEf and guarantees the QoS. In the energy-based cell association scheme, it is crucial to identify the appropriate values for the association bias parameters to optimize the network's CEf.

\section{Energy-based user association scheme}
\subsection{Problem modeling and solution approach}
The analysis in the previous section demonstrated that the CEf can be improved by appropriately adjusting the association bias parameters. In that discussion, we considered a simple functional form where all bias parameters were adjusted using a single parameter. Now, we propose a general optimization framework.
Let $\eta_{\text{ce}}=\frac{\mathcal{R}}{E_{\text{tot}}}=l(\mathbf{B})$, 
where $l:\mathbb{R}^{(T+ 1)\times 1}\rightarrow \mathbb{R}$ is the function between the association bias $\mathbf{B}$ and the CEf $\eta_{\text{ce}}$. The CEf optimization problem is formulated as follows:
\begin{subequations}\label{e33}
\begin{align}
\max_{\mathbf{B}} \quad & \eta_{\text{ce}}=l(\mathbf{B})=\frac{\mathcal{R}}{E_{\text{tot}}},\tag{\ref{e33}}\\
\text {s.t.} \quad& \mathrm{P}_{\text{succ}}(\tau)>\mathrm{P}_{\text{req}},\label{e33a} \tag{\ref{e33}{a}}\\
& B_0=1,\label{e33b} \tag{\ref{e33}{b}}
\end{align}
\end{subequations}
where the constraint (\ref{e33a}) represents the QoS constraint, which stipulates that the probability of a successful transmission must exceed the requisite level for the users $\mathrm{P}_{\text{req}}$.

To solve the optimization problem given by (\ref{e33})-(\ref{e33a}) 
design the energy-based cell association scheme, a genetic algorithm is used. 
The pseudo-code of this algorithm is shown in Algorithm 2.

\begin{algorithm}[tb]
\caption{Genetic algorithm for optimizing the energy-based cell association scheme}\label{alg:alg2}
\begin{algorithmic}
\STATE 
\STATE \textbf{Input:} 
  RE generation rate, $\nu$; 
  user-service rate, $\mu$;  
  energy-consumption rate, $\omega$; 
  number of channels, $N$;  
  density of BSs, $\lambda_B$;
  density of uniformly dist. users, $\lambda_{u(1)}$;
  density of social hotspots, $\lambda_{p}$;
  density of nonuniformly dist. users; $\lambda_{u(2)}$;
  radius of social hotspots, $r$;
  path-loss coefficient, $\alpha$;
  average number of users, $\mathbf{U}$;
  steady-state probabilities of BSs, $\mathbf{\Pi}$;
\STATE \textbf{Output:} Association bias $\mathbf{B}$;
\STATE \hspace{0.5cm}\textbf{Initialization:}Population size, maximum number of iter-
\STATE \hspace{0.5cm}ations, probability of mutation, probability of crossover,
\STATE \hspace{0.5cm}initial population $\mathbf{B}^0$;
\STATE \hspace{0.5cm}\textbf{Repeat:}
\STATE \hspace{1cm} Calculate $l(\mathbf{B}^t)$ based on \eqref{e33}
\STATE \hspace{1cm} Select individuals using roulette algorithm 
\STATE \hspace{1cm} Selected individuals reproduce in pairs 
\STATE \hspace{1cm} New individuals are obtained based on crossover
\STATE \hspace{1cm} and mutation probability.
\STATE \hspace{1cm} Remove individuals with low fitness 
\STATE \hspace{1cm} Form a new population $\mathbf{B}^{t+1}$
\STATE \hspace{1cm} $t=t+1$
\STATE \hspace{0.5cm}\textbf{Until:} $t>T_{\max}$
\STATE \hspace{0.5cm} $\mathbf{B}=\underset{\mathbf{B}_k^{T_{\max}}}{\max}l(\mathbf{B}^{T_{\max}}) $
\end{algorithmic}
\label{alg2}
\end{algorithm}

In Algorithm 2, the chromosome coding of an individual $k$ in iteration $t$ is defined as the association bias of the BSs $\mathbf{B}_k^t\in\mathbb{R}^{(T+ 1)\times 1}$, $k=1,2,\ldots,N_\text{pop}$, where $N_\text{pop}$ is the size of the population. 
The fitness of an individual $k$ in iteration $t$ is $l\left(\mathbf{B}_k^t\right)$. 
Denote by $\mathbf{B}^t=\left(\mathbf{B}_1^t,\mathbf{B}_2^t,\ldots,\mathbf{B}_{N_\text{pop}}^t\right)^\top$, 
$\mathbf{B}^t\in \mathbb{R}^{(T+ 1)\times N_\text{pop}}$, 
the chromosome coding of the population in iteration $t$.

In this algorithm, the individuals to be selected for reproduction are determined on the basis of their fitness. The selection process is based on the roulette algorithm, which assigns a high probability of selection to the individual with the highest fitness value. 
According to the probability of crossover and mutation, the selected individuals are subjected to crossover and mutation in pairs. 
In the event that a crossover operation is carried out, a crossover point is randomly selected from the individual chromosome code. The codes that precede and succeed the crossover point of the two individuals are then exchanged, resulting in the generation of the new individual. Subsequently, a mutation point is randomly selected in the individual chromosomal coding. 
The data associated with this point is then randomly altered to generate the new individual. The fitness of the individuals who underwent cross-over and mutation is recalculated and arranged in descending order. In order to maintain a constant population size, those individuals with lower fitness levels are removed to form the new population. Upon reaching the maximum iteration time, the chromosome coding of the individual with the highest fitness value within the population represents the solution to the problem (\ref{e32}), which is the association bias of the energy-based cell association scheme that maximizes the CEf.

The computational complexity of the genetic algorithm is depended on the size of the population $N_\text{pop}$,and the size of the chromosome of an individual, $T+1$. In the fitness calculation step, the computational complexity is $O(N_\text{pop})$. Using the roulette algorithm, the computational complexity of the selection process is $O(N_\text{pop})$, while the crossover and mutation process has a computational complexity of $O(N_\text{pop}(T+1))$. Therefore, the overall computational complexity of the genetic algorithm is $O(N_\text{pop})+O(N_\text{pop})+O(N_\text{pop}(T+1))+O(N_\text{pop}(T+1))=O(N_\text{pop}(T+1))$. 

\subsection{Simulation results}
We set the size of the population to $N_\text{pop}=50$, the maximum number of iterations to $T_{\max} = 100$, the probability of mutation to 0.2, the probability of crossover to 0.7, the RE-generation rate to $\nu=40$. Based on the results shown in Fig. 3, the maximum achievable probability of successful transmission in the proposed system model is approximately 0.96. And lower values of $\mathrm{P}_{\text{req}}$ is impracticality in real-world scenarios. The requisite level for users is set as $\mathrm{P}_{\text{req}}=0.95$. The other parameters remain the same as those in Table 1.

We denote the energy-based cell association scheme proposed in this section as GA optimized, and denote the energy-aware cell association scheme proposed in \cite{zhang_energy-aware_2017} as '[29]'. 
The association-bias parameter $\beta$ of $B_i=(i+1)^{\beta}$ that is capable of maximizing the CEf of the network and meeting the requisite level for users is selected as 1. 

Next, we make a comparison between these two energy-based cell association schemes with the energy-aware cell association scheme in \cite{zhang_energy-aware_2017} and the traditional nearest-cell association.

\begin{figure}[t]
\centering
\includegraphics[width=3in]{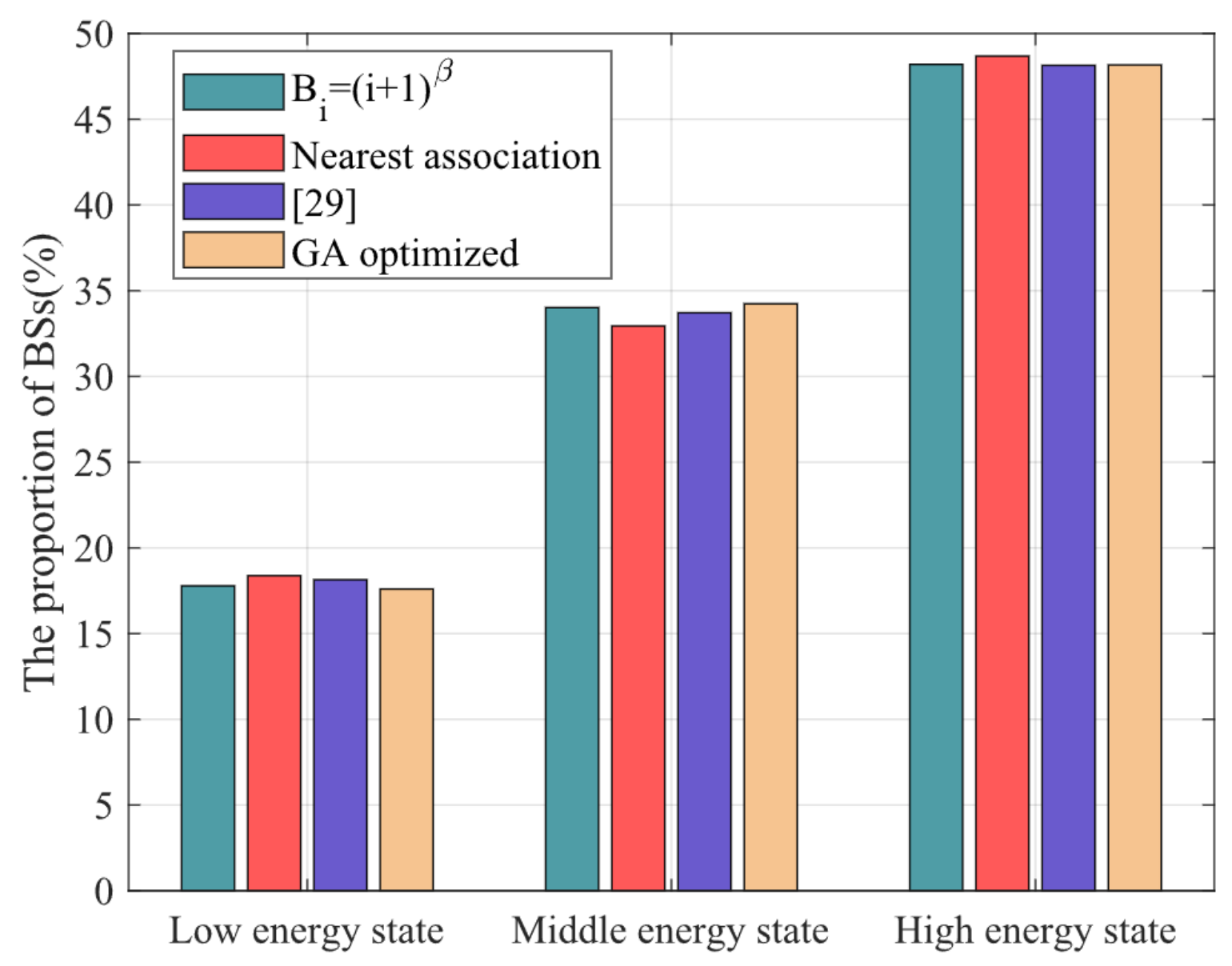}%
\label{fig_7}
\caption{Proportion of BSs in the different battery level categories.} 
\end{figure}

\begin{figure}[t]
\centering
\includegraphics[width=3in]{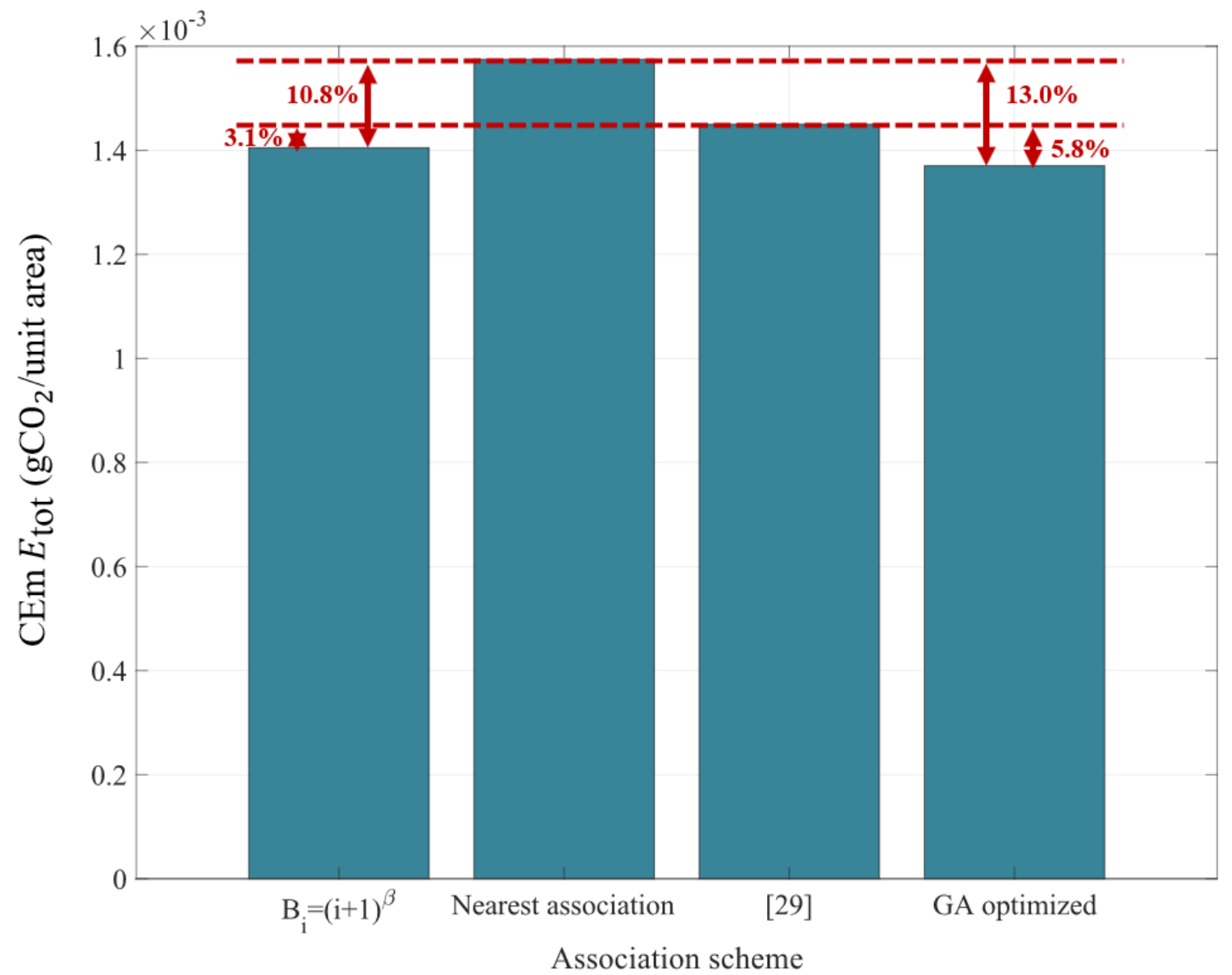}%
\label{fig_8}
\caption{Average CEm per unit area.}
\end{figure}

\begin{figure}[t]
\centering
\includegraphics[width=3in]{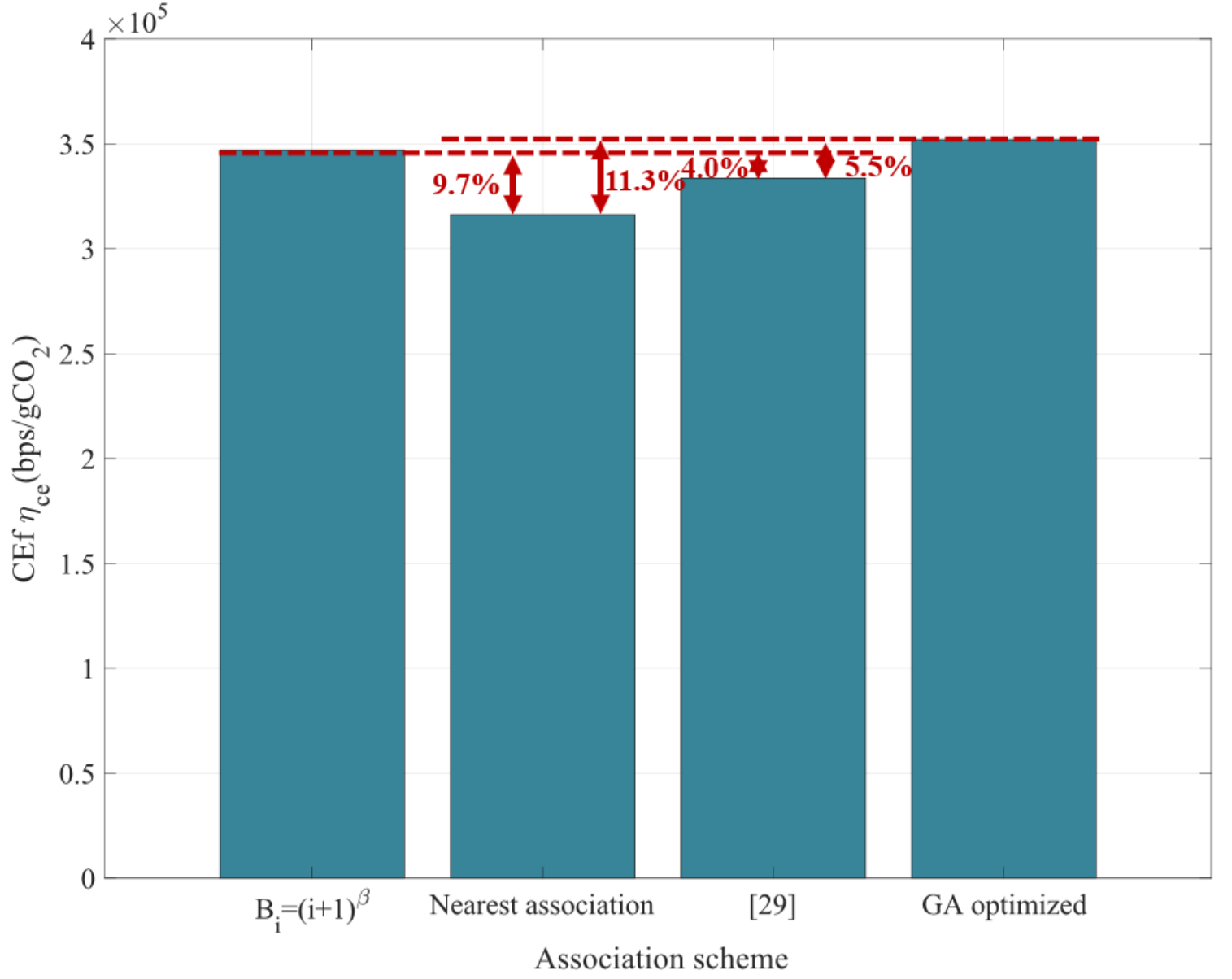}%
\label{fig_9}
\caption{CEf of networks.}
\end{figure}

The proportion of BSs in the different battery level categories for cell association schemes is shown in Fig. 7. The utilization of a nearest-cell association scheme and the energy-aware cell association scheme in \cite{zhang_energy-aware_2017} results in a greater proportion of the BSs with a low-RE state than that observed with the $B_i=(i+1)^\beta$ and GA optimized. Moreover, the proportion of the BSs with a low-RE state is larger when employing the $B_i=(i+1)^\beta$ cell association scheme compared to the GA optimized. It can be concluded that the GA optimized cell association scheme has the potential to reduce the probability of the BSs consuming energy from the grid, and consequently to reduce the CEm of the network.

Fig. 8 depicts the average CEm of the network using different association schemes. The CEm under the traditional nearest-cell association scheme is the largest, followed by the energy-aware cell association scheme in \cite{zhang_energy-aware_2017} and the $B_i=(i+1)^\beta$ cell association scheme. The CEm under the GA optimized cell association scheme is the lowest. The GA optimized cell association scheme efficiently reduces the CEm of the network. Compared with the energy-aware cell association scheme in \cite{zhang_energy-aware_2017}, the $B_i=(i+1)^\beta$ cell association scheme reduces the CEm of networks by $3.1\,\%$. Furthermore, the GA optimized cell association scheme reduces the CEm of the networks by $5.8\,\%$. In comparison with the nearest cell association scheme, the $B_i=(i+1)^\beta$ cell association scheme reduces it by $10.8\,\%$, while the GA optimized cell association scheme achieves a $13.0\,\%$ reduction.

Fig. 9 illustrates the CEf of the networks when using different cell association schemes. As shown in Fig. 9, the CEf is largest when using the GA optimized cell association scheme. The CEf is also higher when using the $B_i=(i+1)^\beta$ cell association scheme compared to the energy-aware cell association scheme in \cite{zhang_energy-aware_2017} and the nearest-cell association scheme. Therefore, using the energy-based cell association scheme can improve the CEf of the network. Compared with the energy-aware cell association scheme in \cite{zhang_energy-aware_2017}, the $B_i=(i+1)^\beta$ cell association scheme demonstrates a $4.0\,\%$ improvement in CEf. This enhancement increases further to $5.5\,\%$ with the GA optimized cell association scheme. Furthermore, compared to the conventional nearest-cell association scheme, the $B_i=(i+1)^\beta$ cell association scheme achieves a $9.7\,\%$ CEf gain, while the GA-optimized attains a maximum improvement of $11.3\,\%$.

\section{Discussion}

The objective of this study is to provide a practical framework for the design of RE-powered cellular networks. Our findings reveal a discrepancy between energy efficiency and CEf, suggesting that the latter is more suitable for optimization in RE-powered networks. Furthermore, the study identifies a trade-off between the network throughput and the CEm, which can be leveraged to maximize the CEf. However, the models employed in this paper necessarily involve simplifications, particularly the assumption of independent BS states and the PCP model the distribution of users. While these simplifications facilitate the analysis, they may not fully capture the complexities of real-world networks. Future work will aim to address these limitations by incorporating inter-BS dependencies, adopting more realistic user distributions, and exploring the application of artificial intelligence-based energy management techniques to further optimize network performance, which may further enhance the trade-off between QoS and sustainability \cite{wang2024generative, zhang2024multi}.

\section{Conclusion}
The development of renewable energy offers a potential solution to the issue of high carbon emissions resulting from the energy consumption of 5G base stations. However, the random nature of renewable energy generation and the non-uniformity of 5G base-station load present challenges in aligning power generation and energy consumption. This hinders the advance of renewable energy powered cellular networks. A reasonable cell association scheme can achieve a match between the energy consumption and the renewable RE state of base stations, thereby enhancing the utilization of renewable energy and reducing the grid source consumption of the networks. In accordance with the Markov chain, the state of the base stations is represented by a quasi-birth-death process. Furthermore, the stochastic geometry theory is employed to ascertain the average successful transmission probability of the users, the carbon emissions, and the carbon efficiency of the network with nonuniformly distributed users. The analysis indicates the existence of an optimal energy-based cell association scheme that can maximize carbon efficiency and reduce carbon emissions of the networks. Moreover, an energy-based cell association scheme has been devised, with its design based on a genetic algorithm. The simulation results demonstrate that this cell association scheme can effectively reduce the carbon emissions of networks and enhance the carbon efficiency of networks by 11.3\%.



\bibliographystyle{IEEEtran}
\bibliography{IEEEabrv,Reference}


 

\begin{IEEEbiography}[{\includegraphics[width=1in,height=1.25in,clip,keepaspectratio]{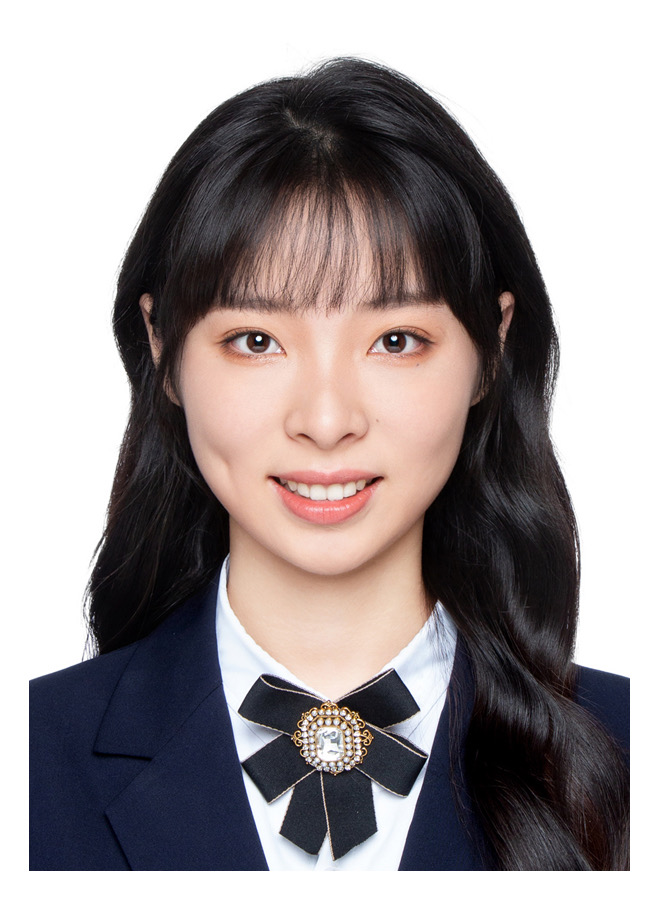}}]{Yuxi Zhao}
is currently working toward the Ph.D. degree with the School of Electronic Information and Communication, Huazhong University of Science and Technology (HUST), Wuhan, China. She received the B.S. degree in communication engineering from HUST, in 2021. Her research interests include green wireless communications, renewable-energy-powered cellular networks. 
\end{IEEEbiography}
\begin{IEEEbiography}[{\includegraphics[width=1in,height=1.25in,clip,keepaspectratio]{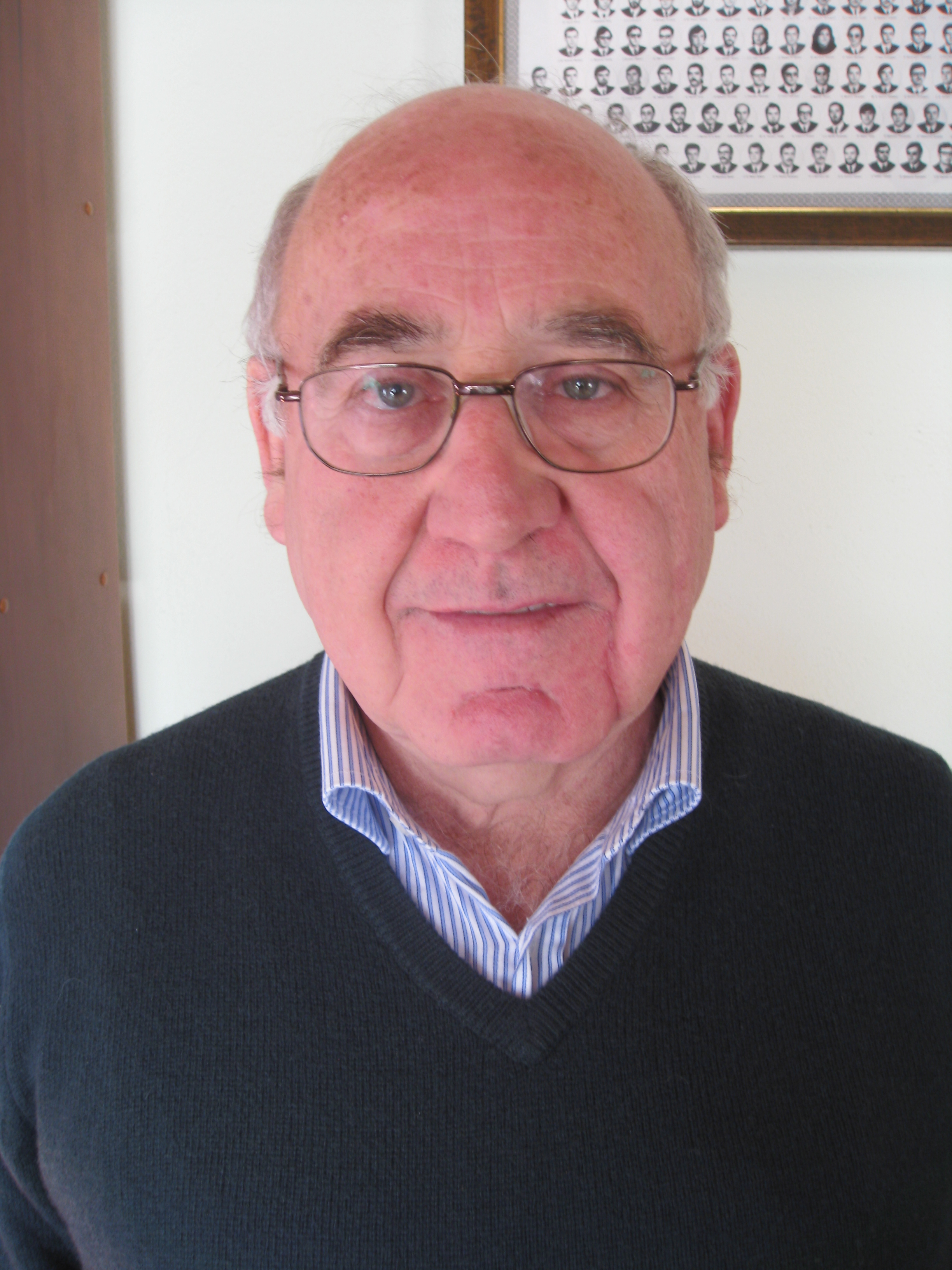}}]{Vicente Casares-Giner}
(Life Member, IEEE) received the telecommunication engineering degree from the Escuela Técnica Superior de Ingenieros en Telecomunicación-Universidad Politécnica de Madrid (ETSIT-UPM) in October 1974 and the Ph.D. degree in telecommunication engineering from the ETSIT-Universidad Politécnica de Catalunya ETSIT-UPC), Barcelona, in September 1980. He was an Assistant Professor in 1974, an Associate Professor in 1985, and a Full Professor in 1991. During the period of 1974–1983, he worked on problems related to signal processing, image restoration, and propagation aspects of radio-link systems. In the first half of 1984, he was a Visiting Scholar with the Royal Institute of Technology (KTH), Stockholm, Sweden, dealing with digital switching and concurrent programming for Stored Program Control (SPC) telephone systems. From September 1994 until August 1995, he was a Visiting Scholar with WINLAB, Rutgers University, USA, working with random access protocols in wireless networks, wireless resource management, and land mobile trunking systems. During the 90’s, he worked in traffic and mobility models in several EU projects. Since September 1996, he has been with the Universitat Politècnica de València (UPV), Valencia, Spain. From 2000 to 2010, he had been involved in multiple Spanish national and EU projects. He was at HUST (Huazhong University of Science and Technology) Wuhan, China, during March 2014 and during March-April 2024. During the first half of 2020, he was a Visiting Professor with the Department of Information and Communication Technology (ICT), University of Agder (UiA), Norway. His research interests include performances evaluation of wireless systems, random access protocols, system capacity and dimensioning, mobility management, cognitive radio, the IoT, and wireless sensor networks. He has served as the General Co-Chair for ISCC 2005 and NGI-2006, and as a TPC Member for many conferences and workshops (Networking 2011, GLOBECOM 2013, ICC 2015, and VTC 2016). In February 2021, he was promoted to Professor Emeritus at UPV.
\end{IEEEbiography}
\begin{IEEEbiography}[{\includegraphics[width=1in,height=1.25in,clip,keepaspectratio]{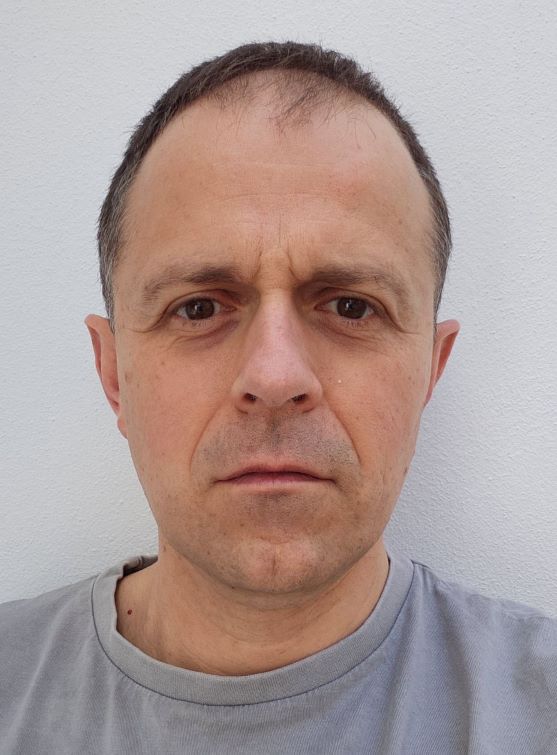}}]{Vicent Pla}
received his Telecommunication Engineering degree (equivalent to a BE+ME) from the Universitat Politècnica de València (UPV), Spain, in 1997 and his PhD in 2005. He also earned a BSc in Mathematics from the Universidad Nacional de Educación a Distancia (UNED), Spain, in 2015. In 1999, he joined UPV's Department of Communications, where he is currently a Professor. His research interests focus on the modeling and performance analysis of communication networks, with an emphasis on traffic and resource management in wireless networks. Dr. Pla has published numerous papers in refereed journals and conference proceedings and has actively participated in several national and European research projects.
\end{IEEEbiography}
\begin{IEEEbiography}[{\includegraphics[width=1in,height=1.25in,clip,keepaspectratio]{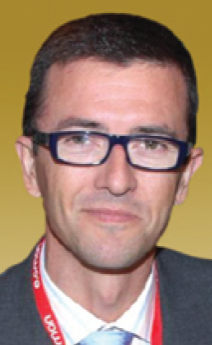}}]{Luis Guijarro}
received the M.Eng. and Ph.D. degrees in telecommunications from the Universitat Politècnica de València (UPV), València, Spain, in 1993 and 1998. He is a Professor in telecommunications economics and regulation with the Department of Communications, UPV. He has coauthored the book Electronic Communications Policy of the European Union (UPV, 2010). He has researched in traffic management in ATM networks and in e-Government and his current research interests include economic modeling of telecommunication service provision. He has published in refereed journals and conferences proceedings in the topics of peer-to-peer interconnection, cognitive radio networks, net neutrality, wireless sensor networks, 5G, 6G, and platform economics.
\end{IEEEbiography}
\begin{IEEEbiography}[{\includegraphics[width=1in,height=1.25in,clip,keepaspectratio]{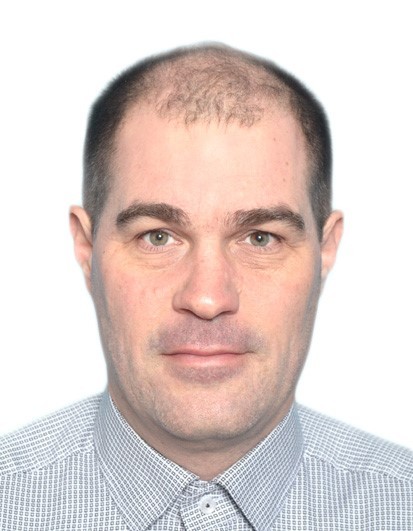}}]{Iztok Humar}
[IEEE SM’10](iztok.humar@fe.uni-lj.si) is a Full Professor and a Vice Dean at the Faculty of Electrical Engineering (FE), University of Ljubljana, Slovenia, EU, where he lecturers on design, management, and modeling of telecommunication networks and fundamentals of electrical engineering. He received Ph.D. in telecommunications from the Faculty of Electrical Engineering (FE) and Ph.D. in information management at the Faculty of Economics, University of Ljubljana, Slovenia, in 2007 and 2009, respectively. His main research topics include the design, planning and management of communications networks and services, and edge cognitive computing and modeling of networks and traffic for energy efficiency and QoS/QoE. He started the collaboration with Huazhong University of Science and Technology, Wuhan, China in 2010 as at a three-month Visiting Professor and Researcher and they continue to perform joint research work through long visiting periods. In Slovenia he served as IEEE Communication Society Chapter Chair and IEEE Section secretary and is a member of Electrical Association of Slovenia.
\end{IEEEbiography}
\begin{IEEEbiography}[{\includegraphics[width=1in,height=1.25in,clip,keepaspectratio]{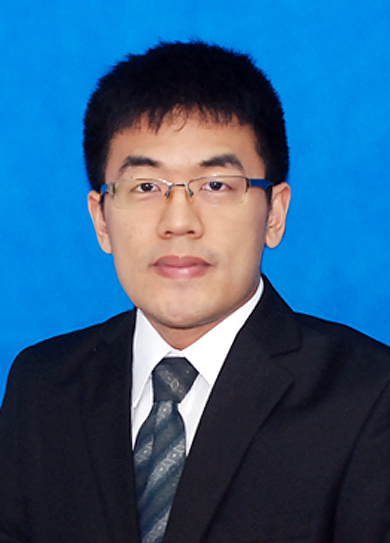}}]{Yi Zhong}
(Senior Member, IEEE) received his B.S. and Ph.D. degrees in Electronic Engineering from the University of Science and Technology of China (USTC) in 2010 and 2015, respectively. From 2015 to 2016, he was a Postdoctoral Research Fellow at the Singapore University of Technology and Design (SUTD), where he worked with the Wireless Networks and Decision Systems (WNDS) Group. He is currently an Associate Professor in the School of Electronic Information and Communications at Huazhong University of Science and Technology (HUST), Wuhan, China. He serves as an associate editor for IEEE Wireless Communications Letters and other academic journals. Dr. Zhong's research interests lie in advanced wireless network theory, with a focus on spatio-temporal modeling, stochastic geometry, spatial network calculus, and energy-efficient communication in next-generation networks. 
\end{IEEEbiography}
\begin{IEEEbiography}[{\includegraphics[width=1in,height=1.25in,clip,keepaspectratio]{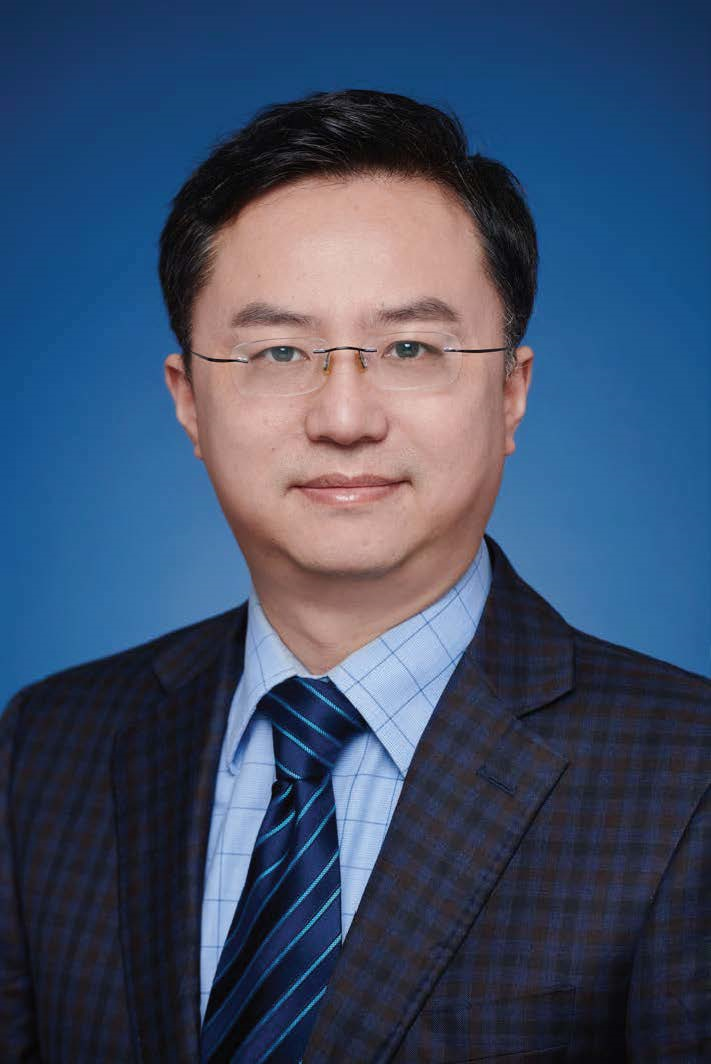}}]{Xiaohu Ge}
(Senior Member, IEEE) is currently a full Professor with the School of Electronic Information and Communications at Huazhong University of Science and Technology (HUST), China. He is an adjunct professor with the Faculty of Engineering and Information Technology at the University of Technology Sydney (UTS), Australia. He received his PhD degree in Communication and Information Engineering from HUST in 2003. He has worked at HUST since Nov. 2005. Prior to that, he worked as a researcher at Ajou University (Korea) and Politecnico Di Torino (Italy) from Jan. 2004 to Oct. 2005. His research interests are in the areas of mobile communications, traffic modeling in wireless networks, green communications, and interference modeling in wireless communications. He has published more than 200 papers in refereed journals and conference proceedings and has been granted about 50 patents in China. He received the Best Paper Awards from IEEE Globecom 2010. Dr. Ge serves as the China Representative for international federation for information processing (IFIP). He serves as an associate editor for IEEE.
\end{IEEEbiography}

\end{document}